\PassOptionsToPackage{square, numbers}{natbib}
\PassOptionsToPackage{table}{xcolor}
\documentclass{article}

\usepackage[preprint]{neurips_2024}

\usepackage[utf8]{inputenc} 
\usepackage[T1]{fontenc}    
\usepackage{hyperref}       
\usepackage{url}            
\usepackage{booktabs}       
\usepackage{amsfonts}       
\usepackage{nicefrac}       
\usepackage{microtype}      
\usepackage[dvipsnames]{xcolor}         
\usepackage{graphicx}
\usepackage{bbding}
\usepackage{fontawesome5}
\usepackage{subcaption,lipsum}
\usepackage{amsmath} 
\usepackage{wrapfig}
\usepackage{enumitem}
\usepackage{xspace}

\usepackage{mdframed}
\newmdenv[
  backgroundcolor=gray!20,
  linecolor=gray!20,
  linewidth=0pt,
  innertopmargin=6pt,
  innerbottommargin=6pt,
  innerleftmargin=6pt,
  innerrightmargin=6pt
]{graybox}

\definecolor{lightcyan}{rgb}{0.7, 1, 1}  

\title{Graph Reasoning for Explainable Cold Start Recommendation}

\author{%
  Jibril Frej \\
  EPFL\\
  \texttt{jibril.frej@epfl.ch} \\
  \And
  Marta Kne\v zevi\'c \\
  EPFL\\
  \texttt{marta.knezevic@epfl.ch} \\
  \And
  Tanja Käser\\
  EPFL\\
  \texttt{tanja.frej@epfl.ch} \\
}

\begin{document}
\newcommand{\grecs}{\texttt{GRECS}\xspace}

\maketitle

\begin{abstract}
The cold start problem, where new users or items have no interaction history, remains a critical challenge in recommender systems (RS).
A common solution involves using Knowledge Graphs (KG) to train entity embeddings or Graph Neural Networks (GNNs). 
Since KGs incorporate auxiliary data and not just user/item interactions, these methods can make relevant recommendations for cold users or items.
Graph Reasoning (GR) methods, however, find paths from users to items to recommend using relations in the KG and, in the context of RS, have been used for interpretability.
In this study, we propose \grecs: a framework for adapting GR to cold start recommendations.
By utilizing explicit paths starting for users rather than relying only on entity embeddings, \grecs can find items corresponding to users' preferences by navigating the graph, even when limited information about users is available.
Our experiments show that \grecs mitigates the cold start problem and outperforms competitive baselines across 5 standard datasets while being explainable. 
This study highlights the potential of GR for developing explainable recommender systems better suited for managing cold users and items.
\end{abstract}

\section{Introduction}
\label{sec:intro}
With the growth of digital commerce and content platforms, users are often overwhelmed by the number of choices available~\cite{iyengar2000choice}, negatively impacting user experience and engagement. 
By recommending a small set of personalized items, recommender systems (RS) can enhance user satisfaction and platform efficiency~\cite{rendle2012bpr,aiolli2013efficient,he2017neural}. 
Moreover, explainable RS can further increase the effectiveness and persuasiveness of an RS \cite{kizilcec2016much, DBLP:journals/corr/abs-1804-11192}. 
Since most of the modern RS uses supervised learning and neural networks (NN) trained on user interactions to make recommendations, two challenges these systems still face are the cold start problem and the lack of explainability of the recommendations. 



The cold start problem occurs when few to no historical interactions are available for new users or items~\cite{schein2002methods,cao2023multi,chaimalas2023bootstrapped,rajapakse2022fast}. 
The challenge extends to both recommending items to users with no or sparse previous history data (cold users) and suggesting new items that have not yet been interacted with by users (cold items). 
Solutions include meta-learning~\cite{wang2022deep,wang2023preference,li2023taml,wu2023m2eu}, using large language models (LLM) to analyze new user preferences/review expressed as natural language~\cite{cao2023multi,sanner2023large}, or using knowledge graphs (KGs) to include additional information about users and items and make recommendations using entity embeddings or graph neural networks (GNNs)~\cite{du2022socially,zheng2018spectral,wang2019multi}. 
However, most of the aforementioned approaches rely on computing a similarity between a user and all items in the dataset, ranking the items by their similarity, and recommending the top-k items. This method of comparing a user to all items makes these approaches more prone to the popularity bias~\cite{bellogin2017statistical} since popular items tend to have a high similarity with users when supervised learning is used. This can result in top recommendations that do not necessarily align with the user's preferences. Furthermore, the majority of the suggested approaches provide no explanations for the recommendations.

The absence of explainability can diminish users' trust and reduce their willingness to follow the recommendation~\cite{kizilcec2016much, DBLP:journals/corr/abs-1804-11192}.
This is especially problematic in high-impact domains such as education where a learner needs to be given a clear explanation when recommended with a course~\cite{kizilcec2016much}. 
Enhancing explainability in NN-based RS often involves using KGs that represent explicit relations between understandable entities~\cite{ai2018learning, fu2020fairness, xie2021explainable}.
Moreover, recent work shows that using graph reasoning (GR) to make a recommendation by finding a path from a user to an item in a KG leads to explainable recommendations while keeping performance competitive to black-box approaches~\cite{xian2019reinforcement,song2019ekar,geng2022path,balloccu2023faithful,frej2024finding}.

In this paper, we leverage GR methods for explainable cold start recommendation. Since GR relies not only on user-item similarity, but also on the structure of the KG to move from a user to an item with a limited number of steps, the set of recommendable items is effectively reduced to reachable items only. We hence hypothesize that GR approaches are less susceptible to the popularity bias and more likely to recommend relevant items in the absence of interaction data.

\textbf{Contributions.} We propose \grecs: a lightweight Graph Reasoning for Explainable Cold Start framework that adapts GR for cold start recommendations, while keeping recommendations explainable. Specifically, we integrate cold entities (users or items) into a KG by employing non-interaction relations and computing relevant embeddings. \grecs can therefore be used on alreay optimized models whithout any additional training. Our implementation is publicly available\footnote{\url{https://anonymous.4open.science/r/cold_rec-B765}}. We evaluate our approach on five publicly available e-commerce and MOOC datasets and demonstrate that \grecs:



\begin{enumerate}[leftmargin=*,label={\textbf{[\arabic*]}}]
    \item provides \textbf{relevant recommendations} to \textbf{strict cold start users} with no prior interactions, outperforming state-of-the-art baseline methods,
    \item recommends relevant \textbf{strict cold start items} with no prior interactions, again outperforming baselines on cold item coverage, 
    \item is \textbf{less sensitive} to the \textbf{popularity bias} than existing methods,
    \item require \textbf{few relations or interactions} to make relevant recommendations.
\end{enumerate}



\vspace{-1em}

\section{Background}

\textbf{Cold Start Recommendation.} A cold start scenario occurs when there is no interaction data available for a user or an item. Common solutions include meta-learning and KG-based approaches.

\vspace{-0.25em}
\textit{Meta-learning} involves learning prior (or meta) knowledge by solving multiple tasks to improve the generalization capacities of models on new examples or new tasks~\cite{wang2022deep}. 
In the context of cold start recommendation, each user is treated as a different task, and the meta-knowledge enablse the RS to make relevant recommendations to new users. 
To further improve performance in the cold start-scenario, PDMA~\cite{wang2023preference} proposes to distinguish new users and novel preferences by decoupling the preference of a cold-start user into common preference transfer and novel preference adaptation. M2eu~\cite{wu2023m2eu} uses similar warm user embeddings to enhance cold start user embeddings, and TAML~\cite{li2023taml} proposes to consider the temporal factor of users' preferences for cold items in the context of news recommendations, where the time factor is crucial to consider.

\vspace{-0.25em}
\textit{KG-based Approaches} use additional information about users and items from a KG to address the cold start problem.
Since user-item interactions are not the only types of information being used in a KG, the approaches can make meaningful recommendations in a cold start scenario. 
MKR~\cite{wang2019multi} trains entity embeddings using multiple tasks over the KG to transfer knowledge between the tasks. This approach produces embeddings that can be used for cold start recommendation, KG completion, or cross-domain recommendation.
UCC~\cite{liu2023uncertainty} generates interactions for cold users that are distributed similarly to those of warm users.
SDCRec~\cite{du2022socially} uses GNN to cold start social recommendation while reducing the popularity bias using contrastive learning.
Metakg~\cite{du2022metakg}, a new framework to use Meta-Learning on KG, combines both approaches for cold start recommendation. Recently, several approaches have used Large Language Models (LLMs) to make cold-start recommendations. ColdGPT~\cite{cao2023multi} proposes to address the strict cold-start item recommendation by using LLMs to find keywords in item descriptions and reviews and include them in a KG to be used for recommendation.~\cite{sanner2023large} propose to tackle the user cold-start problem by analyzing user preferences expressed in natural language with an LLM to make cold-start recommendations.

\vspace{-0.25em}
Unfortunately, the aformentioned methods suffer from the popularity bias in cold start scenarios and the majority of them do not provide explainable recommendations.

\textbf{Graph Reasoning for Explainable Recommendation.}
GR for recommendations identifies paths from a user to the recommended item in a KG, utilizing the relations between entities in the KG to provide human-understandable explanations for recommendations. PGPR~\cite{xian2019reinforcement} was the first approach to use graph reasoning for explainable recommendations, introducing a new Reinforcement Learning (RL) framework for recommendation using GR and pre-defined path patterns. 
Since PGPR, several approaches have been proposed to improve the quality of the recommendations or explanations. 
EKAR~\cite{song2019ekar} is a framework similar to PGPR that uses a different reward to better guide the agent towards items in the training set. 
CAFE~\cite{xian2020cafe} proposes to replace RL with behavioral cloning to avoid some of the shortcomings emanating from RL. 
ReMR~\cite{wang2022multi} proposes to organize the KG into multiple abstraction levels to better reveal users' interests. Finally, UPGPR~\cite{frej2024finding} adopts PGPR to educational use cases and improves its generalizability by removing the pre-defined patterns.

\vspace{-0.25em}
Recent approaches, such as PLM-Rec~\cite{geng2022path} have proposed to move beyond explicit GR by training LLMs over the relations triplets of the KG, avoiding the recall bias caused by unreachable items. While providing better recommendations, the explanations generated by the LLMs are not constrained by the KG, making them less faithful. PEARLM~\cite{balloccu2023faithful} proposes to address this problem by introducing constraints on the sequence decoding to guarantee path faithfulness. 

\vspace{-0.25em}
To the best of our knowledge, \grecs is the first framework for adapting GR for cold start recommendations, resulting in a decreased popularity bias and explainable recommendations.

\vspace{-2mm}
\section{Methodology}
\label{sec:methodology}
\vspace{-1mm}
\grecs, our framework for Graph Reasoning for Explainable Cold Start, adapts GR to cold start scenarios without requiring further training. In the following, we first define the Knowledge Graph Reasoning for Explainable Recommendations (KGRE-Rec) problem and then describe the GR models building the basis for the recommendations. (PGPR~\cite{xian2019reinforcement} and UPGPR~\cite{frej2024finding}). Finally, we detail how \grecs integrates cold entities in the KG and assigns them with meaningful embeddings.

\vspace{-2mm}
\subsection{Problem Formulation}
\begin{figure}
  \centering
    \includegraphics[trim={0 0 0 0},clip,width=\linewidth]{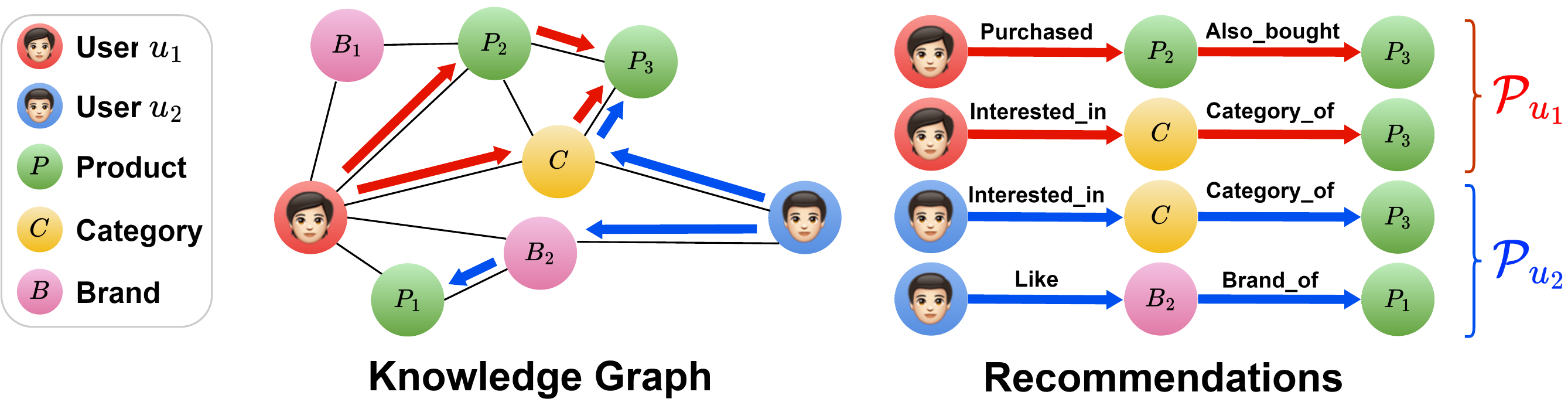}
    \caption{Illustration of the KGRE-REC problem on a KG with a warm user $u_1$ (in red), who has already purchased a product and a cold user $u_2$ (in blue) who has not purchased any product, but whose preferences are assumed to be known (e.g. through a questionnaire). The goal is to find paths from the users to the items to be recommended. The red and blue arrows indicate the paths chosen for the recommendations. While the agent uses past interactions for the warm start user (\textit{purchased}), only non-interaction relations are leveraged to provide a recommendation for the cold start user. All recommendations are explainable as they use entities and relations from the KG.}
  \label{fig:KGRE-Rec}
\vspace{-3mm}
\end{figure}

\vspace{-0.5mm}
 We define a KG as a set of triplets composed of a head entity $e_h$, a tail entity $e_t$, and a relation $r$: $\mathcal{G} = \{(e_h, r, e_t) | e_h, e_t \in \mathcal{E}, r \in \mathcal{R}\}$.
$\mathcal{E}$ and $\mathcal{R}$ denote respectively the entity set and the relations set associated with $\mathcal{G}$. 
We assume that the set of entities contains a set of users $\mathcal{U}$ and items $\mathcal{I}$: $\mathcal{U} \subset \mathcal{E}$ and $\mathcal{I} \subset \mathcal{E}$. 
If a user $u$ has interacted with an item $i$, we add the triplet $(u, interacted ,i)$ to $\mathcal{G}$. 
We define a $k-$hop path in $\mathcal{G}$ as a sequence of $k+1$ entities connected by $k$ relations: $p=\left[e_1, r_1, e_2, r_2, \dots, r_{k}, e_{k+1}\right]$, where
all entities in the path are unique (a path does not use the same entity twice) and all triplets are in the KG: $\forall i \in 1...k: (e_i,r_i,e_{i+1}) \in p \rightarrow (e_i,r_i,e_{i+1}) \in \mathcal{G}$.
Given a user $u \in \mathcal{U}$, KGRE-Rec aims to find a set of paths $\mathcal{P}_u$, each starting from $u$ and terminating at a recommended item $i$, as illustrated in Figure~\ref{fig:KGRE-Rec}.

\vspace{-2mm}
\subsection{Graph Reasoning Models}
\vspace{-1mm}
We use two GR models for explainable recommendations as a basis for \grecs: PGPR and UPGPR.

\vspace{-0.25em}
\textbf{PGPR}. PGPR~\cite{xian2019reinforcement} consists of two main steps: (1) train a KG embedding model to get entities and relations embeddings; (2) optimize an RL agent to find paths in the KG from users to items. We refer the reader to the original paper~\cite{xian2019reinforcement} for a full description of the agent optimization.

The goal of the first step is to embed every entity and relation in $\mathcal{G}$. PGPR uses translational embeddings~\cite{bordes2013translating}, trained by maximizing the conditional probability of each triplet $(e_t, r, e_h) \in \mathcal{G}$:
\begin{equation}
\mathbb{P}(e_t|e_h, r) = \frac{\exp(f(e_h, e_t | r))}{\sum_{e_t^\prime \in E} \exp(f(e_h, e_t^\prime | r))}    
\end{equation}
where $f$ is a translation similarity function between entities $e_h$ and $e_t$ connected with relation $r$:
\begin{equation}
f(e_h, e_t | r) = <\boldsymbol{e_h} + \boldsymbol{r}, \boldsymbol{e_t}> + b_{e_t}
\label{eq:sim}
\end{equation}
where $\boldsymbol{e_h}, \boldsymbol{r}, \boldsymbol{e_t}$ are the embeddings of $e_h, r$ and $e_t$ respectively and $b_{e_t}$ is the bias of $e_t$.

The goal of the second step is to optimize the RL agent, formulating KGRE-Rec as a Markov Decision Process (MDP). The \textit{state} is a vector built by concatenating all entities and relations in the agent's path. 
The initial state is always a user embedding.
For example, if we consider the current path the agent took as $\left[u, r_1, e_1\right]$, the state $s_1$ is defined as $u || r_1 || e_1$, where $||$ is the concatenation operation.
The \textit{action} is defined as the choice of the next hop from the current entity $e_t$.
Hence, the action space $A_t$ consists of all possible relations connected to entity $e_t$ except the ones that lead to already visited graph entities. 
The agent can also take the action \textit{self-loop} that keeps it on the current entity.  
KGRE-Rec has a deterministic \textit{state transition} function, meaning that when that agent takes the action corresponding to going from entity $e_i$ to entity $e_{i+1}$ using relation $r$, the probability of the agent arriving on $e_{i+1}$ for its next state is equal to 1.
Finally, the \textit{reward} is computed using the similarity function $f$ defined in equation~\ref{eq:sim} and using a set of predefined path patterns $\mathcal{P}$:

\begin{equation}
    R(u, e_T, p_T) = \begin{cases}
    \frac{f(u, e_T | r)}{\max_{i \in \mathcal{I} } f(u, i | r)}& \text{if } p_T \in \mathcal{P}\\
    0,              & \text{otherwise}
\end{cases}
\end{equation}
where $e_T$ is the final entity in the path taken by the agent, $r$ is the relation $interacted$, and $\mathcal{P}$ is a set of predefined path patterns. 
We define a $k-$hop path pattern in $\mathcal{G}$ as a sequence of $k+1$ entity types connected by $k$ relations.
Using predefined path patterns in the reward constrains the exploration of the agent toward paths that are relevant for explainability and that end on an item.
However, the manual path definition requires time, expert-knowledge, and can be infeasible in practice for long paths with a large number of possible patterns.

\vspace{-0.25em}
\textbf{UPGPR}.
To address the issues arising with the use of predefined path patterns in the reward, UPGPR~\cite{frej2024finding} uses a new binary reward scheme, allowing to remove the predefined patterns without hurting the performances.
Specifically, the agent receives a reward if it begins at user $u$ and the final entity $e_T$ is an item that $u$ interacted with:
\begin{equation}
  R_{0/1}(u, e_T) = \begin{cases}
    1 &\text{ if } e_T \in \mathcal{I}_u \text{ and } n_{sl} < k-1\\
    0,              & \text{otherwise}
\end{cases}  
\end{equation}
where $\mathcal{I}_u \subseteq I$ is the set of items user $u$ interacted with, $k$ is the number of hops in the path and $n_{sl}$ is the number of self loops. A reward of 0 is assigned to the agent if it uses $k-1$ self-loops because rewarding paths with only one effective hop would lead to poor generalization. For example, if an agent was rewarded for the following path: $\left[user_1,~interacted,~item_1,~\textit{self-loop},~item_1 \right]$, there would be a risk of overfitting to the training interactions. The other components of UPGPR, namely the KG embeddings, \textit{state}, \textit{actions}, \textit{state transitions}, and \textit{agent optimization} are the same as PGPR.

\vspace{-1mm}
\subsection{Graph Reasoning for Explainable Cold Start Recommendation}
Our \grecs framework adapts GR models for cold start recommendations after the embeddings and agent are optimized. Hence, \grecs can be used on an already optimized model, not requiring any additional training. Specifically, \grecs follows two steps: (1) integrating the new entity (user or item) into the KG using their known (non-interaction) relations; (2) assigning the entity with a meaningful embedding using its neighboring entities' embeddings.

\vspace{-0.25em}
\textbf{KG integration.}  Given a new entity $e \notin \mathcal{E}$ related to a small set of entities via non-interaction relations ($r \neq interacted$), we add all triplets containing $e$ to $\mathcal{G}$. Thanks to this integration, the agent will have the capacity to either start from a cold user or land on a cold item using the non-interaction relations connected to these entities that were added to the graph. It should be noted that we are assuming to know at least one relation of the cold entity to integrate. For example, a new user can indicate the brand(s) they like on a questionnaire. If a new entity is not related to any existing entity in the KG, it will not be used by \grecs.

\textbf{Cold Embeddings.} While the agent can navigate the KG from a cold user (or to a cold item) via their integration in the KG, it needs meaningful embeddings in its state representation to take an action that will lead to a relevant recommendation. To this end, we propose to calculate the embedding for a new entity by using the average translations from its related entities:
\begin{equation}
\boldsymbol{e} = \sum_{(r', e'_t) \in \mathcal{G}_{e}} \left(\boldsymbol{e'_t} - \boldsymbol{r'}\right)/|\mathcal{G}_{e}|
\label{eq:cold_emb}
\end{equation}
where $\mathcal{G}_{e}$ is the subset of all triplets in $\mathcal{G}$ whose head entity is $e$.
This choice is motivated by the KG embeddings being trained using a translation method as described in equation~(\ref{eq:sim}). To evaluate our cold embeddings assignment strategy, we will also compare it to using null embeddings (zeros values everywhere) that correspond to no prior knowledge about users or items. In the following sections, we denote models using the average translation embeddings as PGPR$_{a}$/UPGPR$_{a}$, null embeddings as PGPR$_0$/UPGPR$_0$, and both methods regardless of the embeddings as PGPR/UPGPR.

\vspace{-2mm}
\section{Datasets}
\label{sec:experiments}
\vspace{-2mm}
We evaluated \grecs on four e-commerce datasets~\cite{he2016ups} and one education dataset~\cite{dessi2018coco}.

\textbf{E-commerce Datasets.} The four e-commerce datasets~\cite{he2016ups} are: Amazon's \textit{Beauty}, \textit{CDs and Vinyl}, \textit{Clothing}, and \textit{Cellphones}.\footnote{Datasets can be downloaded~\href{https://drive.usercontent.google.com/download?id=1CL4Pjumj9d7fUDQb1_leIMOot73kVxKB}{\color{blue}{here}}.}
These datasets consist of user purchases, product reviews, and meta-information about products. 
In the context of the e-commerce datasets, we are interested in recommending a \textit{Product} to a \textit{User} and we use the relation \textit{purchase} as the interaction. 
The full list of entities, relations, and statistics of the e-commerce datasets are displayed in Table~\ref{table:dataset_statistics} in Appendix~\ref{sec:appendix_datasets}. 
To integrate cold start users in the KG, we added 2 relations to the original KG: \textit{interested\_in} and \textit{like}. We connect a \textit{User} with a \textit{Category} using the relation \textit{interested\_in} if the user purchased a product belonging to that category. Similarly, we connect a \textit{User} with a \textit{Brand} using the relation \textit{like} if the user purchased a product belonging to that brand.

\textbf{Education Dataset.} COCO~\cite{dessi2018coco} consists of online course from the Udemy\footnote{\url{https://www.udemy.com/.}} platform. 
This dataset provides user enrollments, course descriptions, instructors, and categories.
On the COCO dataset, we recommend a \textit{Course} to a \textit{User} using the \textit{enroll} relation as the interaction.
The full list of entities, relations, and statistics of the dataset are in Table~\ref{table:dataset_statistics} in Appendix~\ref{sec:appendix_datasets}. 
To include cold start users in the KG, we added the entity \textit{Skill} and the relations \textit{know} and \textit{cover} to the original KG. Skills taught in each course were extracted from the course description using skillNER\footnote{\url{https://github.com/AnasAito/SkillNER}}. 
If a \textit{Skill} is mentioned in a \textit{Course} description, we connect them in the KG using the relation \textit{cover}. If a user enrolled in a \textit{Course} with the relation \textit{cover} with a \textit{Skill}, we connect the \textit{User} and the \textit{Skill} using the relation \textit{know}.

\textbf{Datasets Splits.}
All datasets were split into training, validation, and test sets at the user level. 
For each user, the first $80\%$ of their interactions (\textit{purchase} for the e-commerce datasets and \textit{enroll} for COCO) were selected to be in the training set, the next $10\%$ were added to the validation set, and the last $10\%$ were added to the test set.
Additionally, we randomly selected $20\%$ of users and items to be strict cold start users and items and removed them from the training set. Half of the strict cold start users were assigned to the validation set and the other half to the test set. When adding them to the validation and test sets, we integrated cold start users into the e-commerce KG using the relations \textit{interested\_in} and \textit{like} and into the COCO KG using the relation \textit{has\_skills}.
Similarly, cold start items were integrated into the e-commerce KG using the relations \textit{belong\_to} and \textit{produced\_by} and into the COCO KG using the relation \textit{cover}.
To study the impact of the number of known relations and to have realistic cold start users, we limited the occurrences of the relations in the validation and the test KG. 
Specifically, we uniformly sampled a random integer $k$ from 1 to 10 for each cold start user and relation and connected the user only to the top $k$ entities in terms of relation frequency. For example, on the e-commerce datasets, this is similar to asking a new user to indicate their $k$ favorite brands, and including the user using the \textit{like} relation with their favorite brands. 



\section{Results}
\label{sec:results}
We performed four experiments to demonstrate the advantages of \grecs. \textbf{[1]} \grecs outperforms all baselines on cold start users in terms of nDCG and HR. \textbf{[2]} \grecs 
 outperforms all baselines in cold item coverage. \textbf{[3]} \grecs mitigates popularity bias. \textbf{[4]} \grecs needs only a small number of relations (not interactions) to provide an accurate recommendation, and providing few interactions increases the relevance of recommendations even further. Note that while being adopted for the cold start scenario, GR methods also match or outperform all the baselines on 4/5 datasets on warm users who already interacted with items. We report the full detailed results in Table~\ref{tab:all_results} in Appendix~\ref{sec:appendix_all}.  

\textbf{Baselines.} We compared \grecs to eight competitive baselines with implementations publicly available on Recbole~\cite{zhao2021recbole}.
Pop ranks items based on their popularity, defined by their frequency in the training set.
ItemKNN~\cite{aiolli2013efficient} uses k-nearest-neighbor to recommend items similar to those a user has previously selected.
BPR~\cite{rendle2012bpr} optimizes user-item similarity using a ranking-based Bayesian method.
NeuMF~\cite{he2017neural} combines matrix factorization and a multi-layer perceptron to compute a user-item matching score. 
CFKG~\cite{ai2018learning} uses the similarity between users and items KG embeddings for recommendations and constructs a path from users to items to generate a post-hoc explanation.
KGCN~\cite{wang2019knowledge} uses Knowledge Graph Convolutional Networks (KGCN) to train an end-to-end model for recommendation, directly on the KG.
SpectralCF$^\text{\color{NavyBlue}\tiny\SnowflakeChevron}$~\cite{zheng2018spectral} proposes to address the cold start problem by structuring the interactions between users and items as a bipartite graph and using information in the spectral domains to make relevant recommendations.
MKR$^\text{\color{NavyBlue}\tiny\SnowflakeChevron}$~\cite{wang2019multi} addresses the cold start problem by a multi-task approach that optimizes a model jointly on KG embeddings and for recommendation.

\textbf{Evaluation Metrics.} To assess recommendation relevance, we used two standard ranking metrics: nDCG@$K$ and Hit Ratio@$K$. 
We use the POPB@$K$~\cite{borges2021mitigating} to measure the popularity bias. To evaluate the ability of the models to recommend strict cold items, we adapt 2 metrics from previous literature on long tail recommendation~\cite{abdollahpouri2017controlling,abdollahpouri2019managing}: \textbf{(1)} the Cold Item Coverage (Cov.$^\text{\color{NavyBlue}\tiny\SnowflakeChevron}$) reports the proportion of recommended cold items among all cold items across all users in the test set; \textbf{(2)} the Average Cold Item Proportion@$K$ (Prop.$^\text{\color{NavyBlue}\tiny\SnowflakeChevron}$) reports the average proportion of item recommended for each user.
All metrics are computed on the top 10 recommended items for each user ($K=10$). We define the Long Tail items as the 80\% of items with the lowest number of interactions in the training set. 
We report the average and standard deviation of the metric across 3 runs with different seeds.

\textbf{Implementation.}
For \grecs, we set PGPR and UPGPR embedding dimensions to 100 and trained them for 30 epochs. Other hyperparameters (learning rate, batch size, etc...) were the same as in the original PGPR implementation\footnote{\url{https://github.com/orcax/PGPR}}.
We used the full path history for the agent's state with a path length fixed to 3, making our state composed of six KG entities embedding, each of size 100.
RL agents were trained for 50 epochs using the Adam optimizer~\cite{kingma2014adam} with a learning rate of 0.001.
Baselines were implemented and evaluated with Recbole\footnote{\url{https://recbole.io/}}~\cite{zhao2021recbole} with batches of size 512 and Adam optimizer~\cite{kingma2014adam} with a learning rate of 0.001 for 50 epochs with early stopping on the validation set with a patience of 10 epochs.
Baselines with embeddings had their dimensions set to 100, and all other hyperparameters were set to their default values.
Training and evaluation took 2-4 hours on a single Intel Xeon Gold 6132 CPU on all datasets except CDs (the largest dataset), which took up to 48 hours per model.

\subsection{Exp. 1: \grecs outperforms baselines on strict cold start users}
\label{subsec:C1}

Table~\ref{tab:all_cold_results} presents the performance of \grecs and baselines in a strict cold start scenario across all datasets using user-centric metrics. We observe that:
\textbf{(1)} \grecs outperforms all baselines across all datasets, showing the ability of GR to make recommendations to users with no prior interactions that are more relevant than baselines that rely on heuristics (Pop), matrix factorization (NeuMF) or entity embeddings (CFKG).
\textbf{(2)} Neither of the two cold embedding strategies shows a definitive advantage over the other. 
For instance, null embeddings prove more effective on the Beauty and Cellphones datasets, while the average translation strategy performs better on the Clothing dataset. 
\textbf{(3)} Despite being designed for cold start recommendations, MKR$^\text{\color{NavyBlue}\tiny\SnowflakeChevron}$ and SpectralCF$^\text{\color{NavyBlue}\tiny\SnowflakeChevron}$ underperform against the popularity-based recommender, which does not make personalized recommendations nor use the KG. This highlights the challenge of developing personalized recommendation models that outperform simple heuristic approaches for strict cold-start users.
\textbf{(4)} UPGPR performs better than PGPR on all E-commerce datasets and both methods display similar performances on the COCO dataset.

\begin{table*}[h]
\centering
\resizebox{\textwidth}{!}{%
\begin{tabular}{l|cc|cc|cc|cc|cc}
\multicolumn{1}{c}{} & \multicolumn{2}{c}{\textbf{Beauty}} & \multicolumn{2}{c}{\textbf{CDs}} & \multicolumn{2}{c}{\textbf{Cellphones}} & \multicolumn{2}{c}{\textbf{Clothing}} & \multicolumn{2}{c}{\textbf{COCO}}\\
Measures (\%) & nDCG  & HR & nDCG  & HR & nDCG  & HR & nDCG  & HR & nDCG  & HR \\
\toprule
Pop & \underline{1.03} \footnotesize{$\pm$ 0.0} & \underline{1.92} \footnotesize{$\pm$ 0.0} & 0.28 \footnotesize{$\pm$ 0.0} & 0.76 \footnotesize{$\pm$ 0.0} & \underline{1.67} \footnotesize{$\pm$ 0.0} & \underline{3.24} \footnotesize{$\pm$ 0.0} & \underline{0.48} \footnotesize{$\pm$ 0.0} & 0.87 \footnotesize{$\pm$ 0.0} & \underline{1.33} \footnotesize{$\pm$ 0.0} & 2.61 \footnotesize{$\pm$ 0.0} \\
ItemKNN & 0.04 \footnotesize{$\pm$ 0.0} & 0.10 \footnotesize{$\pm$ 0.0} & 0.00 \footnotesize{$\pm$ 0.0} & 0.01 \footnotesize{$\pm$ 0.0} & 0.01 \footnotesize{$\pm$ 0.0} & 0.04 \footnotesize{$\pm$ 0.0} & 0.00 \footnotesize{$\pm$ 0.0} & 0.00 \footnotesize{$\pm$ 0.0} & 0.00 \footnotesize{$\pm$ 0.0} & 0.01 \footnotesize{$\pm$ 0.0} \\
BPR & 0.18 \footnotesize{$\pm$ 0.0} & 0.40 \footnotesize{$\pm$ 0.1} & 0.01 \footnotesize{$\pm$ 0.0} & 0.03 \footnotesize{$\pm$ 0.0} & 0.36 \footnotesize{$\pm$ 0.2} & 0.68 \footnotesize{$\pm$ 0.3} & 0.10 \footnotesize{$\pm$ 0.0} & 0.19 \footnotesize{$\pm$ 0.1} & 0.07 \footnotesize{$\pm$ 0.0} & 0.16 \footnotesize{$\pm$ 0.0} \\
NeuMF & 0.80 \footnotesize{$\pm$ 0.1} & 1.67 \footnotesize{$\pm$ 0.0} & 0.24 \footnotesize{$\pm$ 0.0} & 0.63 \footnotesize{$\pm$ 0.0} & 1.63 \footnotesize{$\pm$ 0.0} & 3.23 \footnotesize{$\pm$ 0.0} & \underline{0.48} \footnotesize{$\pm$ 0.0} & \underline{0.88} \footnotesize{$\pm$ 0.0} & 0.26 \footnotesize{$\pm$ 0.0} & 0.78 \footnotesize{$\pm$ 0.2} \\
CFKG & 0.69 \footnotesize{$\pm$ 0.0} & 1.48 \footnotesize{$\pm$ 0.1} & \underline{0.30} \footnotesize{$\pm$ 0.0} & \underline{0.78} \footnotesize{$\pm$ 0.1} & 0.64 \footnotesize{$\pm$ 0.1} & 1.50 \footnotesize{$\pm$ 0.3} & 0.17 \footnotesize{$\pm$ 0.0} & 0.33 \footnotesize{$\pm$ 0.1} & 1.01 \footnotesize{$\pm$ 0.4} & 2.32 \footnotesize{$\pm$ 0.7} \\
KGCN & 0.04 \footnotesize{$\pm$ 0.0} & 0.10 \footnotesize{$\pm$ 0.0} & 0.02 \footnotesize{$\pm$ 0.0} & 0.08 \footnotesize{$\pm$ 0.0}  & 0.10 \footnotesize{$\pm$ 0.0} & 0.21 \footnotesize{$\pm$ 0.0} & 0.03 \footnotesize{$\pm$ 0.0} & 0.07 \footnotesize{$\pm$ 0.0} & 0.11 \footnotesize{$\pm$ 0.1} & 0.24 \footnotesize{$\pm$ 0.1} \\
MKR$^\text{\color{NavyBlue}\tiny\SnowflakeChevron}$ & 0.74 \footnotesize{$\pm$ 0.1} & 1.46 \footnotesize{$\pm$ 0.2} & 0.17 \footnotesize{$\pm$ 0.0} & 0.39 \footnotesize{$\pm$ 0.0} & 1.46 \footnotesize{$\pm$ 0.1} & 3.02 \footnotesize{$\pm$ 0.2} & 0.38 \footnotesize{$\pm$ 0.0} & 0.83 \footnotesize{$\pm$ 0.1} & 0.53 \footnotesize{$\pm$ 0.2} & 1.46 \footnotesize{$\pm$ 0.6} \\
SpectralCF$^\text{\color{NavyBlue}\tiny\SnowflakeChevron}$ & 0.81 \footnotesize{$\pm$ 0.2} & 1.76 \footnotesize{$\pm$ 0.4} & 0.21 \footnotesize{$\pm$ 0.0} & 0.54 \footnotesize{$\pm$ 0.1} & 1.54 \footnotesize{$\pm$ 0.1} & 3.17 \footnotesize{$\pm$ 0.1} & 0.45 \footnotesize{$\pm$ 0.1} & 0.80 \footnotesize{$\pm$ 0.2} & 1.27 \footnotesize{$\pm$ 0.4} & \underline{2.63} \footnotesize{$\pm$ 0.8} \\
\midrule
PGPR$_a$ & 1.65 \footnotesize{$\pm$ 0.3} & 4.30 \footnotesize{$\pm$ 0.8} &  0.83 \footnotesize{$\pm$ 0.1} & 1.49 \footnotesize{$\pm$ 0.2}  & 0.78 \footnotesize{$\pm$ 0.2} & 1.77 \footnotesize{$\pm$ 0.3} & 1.07 \footnotesize{$\pm$ 0.1} & 2.55 \footnotesize{$\pm$ 0.3} & 2.98 \footnotesize{$\pm$ 0.5} & 4.36 \footnotesize{$\pm$ 0.8} \\
PGPR$_0$ & 1.78 \footnotesize{$\pm$ 0.3} & 3.83 \footnotesize{$\pm$ 0.7} & 0.31 \footnotesize{$\pm$ 0.1} & 0.86 \footnotesize{$\pm$ 0.3} & 1.85 \footnotesize{$\pm$ 0.3} & 4.18 \footnotesize{$\pm$ 0.6} & 0.51 \footnotesize{$\pm$ 0.0} & 1.32 \footnotesize{$\pm$ 0.1} & 3.07 \footnotesize{$\pm$ 0.2} & \textbf{6.83} \footnotesize{$\pm$ 0.3} \\
UPGPR$_a$ & 2.26 \footnotesize{$\pm$ 0.5} & 5.01 \footnotesize{$\pm$ 1.3} & 1.05 \footnotesize{$\pm$ 0.1} & 1.68 \footnotesize{$\pm$ 0.2} & 1.74 \footnotesize{$\pm$ 0.4} & 4.32 \footnotesize{$\pm$ 0.8} & \textbf{1.76} \footnotesize{$\pm$ 0.1} & \textbf{4.05} \footnotesize{$\pm$ 0.2} & \textbf{3.49} \footnotesize{$\pm$ 0.2} & 4.71 \footnotesize{$\pm$ 0.1} \\
UPGPR$_0$ & \textbf{3.07} \footnotesize{$\pm$ 0.2} & \textbf{6.75} \footnotesize{$\pm$ 0.7} & \textbf{1.06} \footnotesize{$\pm$ 0.2} & \textbf{2.61} \footnotesize{$\pm$ 0.4} & \textbf{3.61} \footnotesize{$\pm$ 0.1} & \textbf{7.97} \footnotesize{$\pm$ 0.2} & 1.05 \footnotesize{$\pm$ 0.1} & 2.64 \footnotesize{$\pm$ 0.2} & 1.86 \footnotesize{$\pm$ 0.2} & 3.56 \footnotesize{$\pm$ 0.5} \\
\bottomrule
\end{tabular}
}
\caption{Performance of \grecs compared to the baselines for strict cold start users. The best results are highlighted in bold and the best baseline is underlined.}
\label{tab:all_cold_results}
\end{table*}

To further explore performance differences between PGPR and UPGPR, we analyzed the proportion of paths the agents followed on each dataset.
Table~\ref{table:small_beauty_patterns} shows the proportion of paths on the Beauty test set. The symbol ($^\text{\color{OrangeRed}\tiny\faHotjar}$) represents paths used for warm users and the symbol ($^\text{\color{NavyBlue}\tiny\SnowflakeChevron}$) for cold users. UPGPR selects patterns more evenly than PGPR and frequently uses those suitable for cold start users even in warm start scenarios. Specifically, PGPR uses cold start-compatible patterns (marked in blue in Table ~\ref{table:small_beauty_patterns})  on warm users $19.03\%$ of the time, whereas UPGPR does so $52.70\%$ of the time. This likely contributes to UPGPR's superior performance in cold start situations compared to PGPR since UPGPR's agents are more familiar with patterns that can be used for cold start users. Similarly, on the COCO dataset (see Table~\ref{table:small_coco_patterns}), \grecs applies cold start-compatible patterns to warm users over $90\%$ of the time for both PGPR and UPGPR, explaining their close performances. This trend is consistent across all datasets, with detailed proportions listed in Tables~\ref{table:beauty_patterns},~\ref{table:cds_patterns},~\ref{table:cell_patterns},~\ref{table:cloth_patterns}, and~\ref{table:coco_patterns} in Appendix~\ref{sec:appendix_patterns}.

\begin{table*}[h]
\centering
\resizebox{\textwidth}{!}{
\begin{tabular}{lcccc}
 & \textbf{PGPR$_a^\text{\color{OrangeRed}\tiny\faHotjar}$} & \textbf{UPGPR$_a^\text{\color{OrangeRed}\tiny\faHotjar}$} & \textbf{PGPR$_a^\text{\color{NavyBlue}\tiny\SnowflakeChevron}$} & \textbf{UPGPR$_a^\text{\color{NavyBlue}\tiny\SnowflakeChevron}$} \\
 \midrule
User $\xrightarrow{\text{mentioned}}$ Word $\xrightarrow{\text{described}}$ Product & 71.58\% &  46.72\% & 0\% & 0\%\\
\rowcolor{lightcyan}
User $\xrightarrow{\text{interested\_in}^\text{\color{NavyBlue}\SnowflakeChevron}}$ Category $\xrightarrow{\text{belong\_to}^\text{\color{NavyBlue}\SnowflakeChevron}}$ Product & 16.78\% & 29.18\% & 65.18\% & 64.95\%\\
\rowcolor{lightcyan}
User $\xrightarrow{\text{like}^\text{\color{NavyBlue}\SnowflakeChevron}}$ Brand $\xrightarrow{\text{produced\_by}^\text{\color{NavyBlue}\SnowflakeChevron}}$ Product & 2.26\% & 23.52\% & 31.3\% & 35.0\%\\
\bottomrule
\end{tabular}
}
\caption{Proportion of the three most frequent patterns used by the agents on the test set of the Beauty dataset for warm users ($^\text{\color{OrangeRed}\tiny\faHotjar}$) and cold users ($^\text{\color{NavyBlue}\tiny\SnowflakeChevron}$). Rows highlighted in blue denote cold start-compatible patterns that can be used for recommending an item to a strict cold user.}
\label{table:small_beauty_patterns}
\vspace{-3mm}
\end{table*}

\begin{table*}[h]
\centering
\resizebox{\textwidth}{!}{%
\begin{tabular}{lcccc}
 & \textbf{PGPR$_a^\text{\color{OrangeRed}\tiny\faHotjar}$} & \textbf{UPGPR$_a^\text{\color{OrangeRed}\tiny\faHotjar}$} & \textbf{PGPR$_a^\text{\color{NavyBlue}\tiny\SnowflakeChevron}$} & \textbf{UPGPR$_a^\text{\color{NavyBlue}\tiny\SnowflakeChevron}$} \\
 \midrule
 \rowcolor{lightcyan}
User $\xrightarrow{\text{know}^\text{\color{NavyBlue}\SnowflakeChevron}}$ Skill $\xrightarrow{\text{know}^\text{\color{NavyBlue}\SnowflakeChevron}}$ User $\xrightarrow{\text{enroll}}$ Course & 59.0\% & 9.86\% & 65.45\% & 6.5\%\\
\rowcolor{lightcyan}
User $\xrightarrow{\text{know}^\text{\color{NavyBlue}\SnowflakeChevron}}$ Skill $\xrightarrow{\text{cover}^\text{\color{NavyBlue}\SnowflakeChevron}}$ Course & 40.18\% & 83.84\% & 34.55\%  & 93.5\% \\
User $\xrightarrow{\text{enroll}}$ Course $\xrightarrow{\text{enroll}}$ User $\xrightarrow{\text{enroll}}$ Course & 0.0\% & 5.61\% & 0\% & 0\%\\
\bottomrule
\end{tabular}
}
\caption{Proportion of the three most frequent patterns used by the agents on the test set of the COCO dataset for warm users ($^\text{\color{OrangeRed}\tiny\faHotjar}$) and cold users ($^\text{\color{NavyBlue}\tiny\SnowflakeChevron}$). Rows highlighted in blue denote patterns that can be used for recommending an item to a strict cold user.}
\label{table:small_coco_patterns}
\vspace{-3mm}
\end{table*}

\begin{graybox} \grecs provides the most relevant recommendations for strict cold start users across all 5 datasets.\end{graybox}

\vspace{-1mm}
\subsection{Exp. 2: \grecs achieves high coverage of cold start items}
\label{subsec:C2}
\vspace{-1mm}

\begin{table*}[h]
\centering
\resizebox{\textwidth}{!}{%
\begin{tabular}{l|cc|cc|cc|cc|cc}
\multicolumn{1}{c}{} & \multicolumn{2}{c}{\textbf{Beauty}} & \multicolumn{2}{c}{\textbf{CDs}} & \multicolumn{2}{c}{\textbf{Cellphones}} & \multicolumn{2}{c}{\textbf{Clothing}} & \multicolumn{2}{c}{\textbf{COCO}}\\
Measures (\%) & Cov.$^\text{\color{NavyBlue}\tiny\SnowflakeChevron}$ & Prop.$^\text{\color{NavyBlue}\tiny\SnowflakeChevron}$ & Cov.$^\text{\color{NavyBlue}\tiny\SnowflakeChevron}$ & Prop.$^\text{\color{NavyBlue}\tiny\SnowflakeChevron}$ & Cov.$^\text{\color{NavyBlue}\tiny\SnowflakeChevron}$ & Prop.$^\text{\color{NavyBlue}\tiny\SnowflakeChevron}$ & Cov.$^\text{\color{NavyBlue}\tiny\SnowflakeChevron}$ & Prop.$^\text{\color{NavyBlue}\tiny\SnowflakeChevron}$ & Cov.$^\text{\color{NavyBlue}\tiny\SnowflakeChevron}$ & Prop.$^\text{\color{NavyBlue}\tiny\SnowflakeChevron}$\\
\toprule
Pop & 0.00 \footnotesize{$\pm$ 0.0} & 0.00 \footnotesize{$\pm$ 0.0} & 0.00 \footnotesize{$\pm$ 0.0} & 0.00 \footnotesize{$\pm$ 0.0} & 0.00 \footnotesize{$\pm$ 0.0} & 0.00 \footnotesize{$\pm$ 0.0} & 0.00 \footnotesize{$\pm$ 0.0} & 0.00 \footnotesize{$\pm$ 0.0} & 0.00 \footnotesize{$\pm$ 0.0} & 0.00 \footnotesize{$\pm$ 0.0} \\
ItemKNN & 0.21 \footnotesize{$\pm$ 0.0} & \underline{5.74} \footnotesize{$\pm$ 0.0} & 0.08 \footnotesize{$\pm$ 0.0} & \underline{\textbf{11.34}} \footnotesize{$\pm$ 0.0} & 0.48 \footnotesize{$\pm$ 0.0} & \underline{11.54} \footnotesize{$\pm$ 0.0} & \underline{0.22} \footnotesize{$\pm$ 0.0} & \underline{\textbf{11.51}} \footnotesize{$\pm$ 0.0} & 0.25 \footnotesize{$\pm$ 0.0} & \underline{\textbf{11.42}} \footnotesize{$\pm$ 0.0} \\
BPR & 0.00 \footnotesize{$\pm$ 0.0} & 0.00 \footnotesize{$\pm$ 0.0} & 0.00 \footnotesize{$\pm$ 0.0} & 0.00 \footnotesize{$\pm$ 0.0} & 0.00 \footnotesize{$\pm$ 0.0} & 0.00 \footnotesize{$\pm$ 0.0} & 0.00 \footnotesize{$\pm$ 0.0} & 0.00 \footnotesize{$\pm$ 0.0} & 0.00 \footnotesize{$\pm$ 0.0} & 0.00 \footnotesize{$\pm$ 0.0} \\
NeuMF & 0.00 \footnotesize{$\pm$ 0.0} & 0.00 \footnotesize{$\pm$ 0.0} & 0.00 \footnotesize{$\pm$ 0.0} & 0.00 \footnotesize{$\pm$ 0.0} & 0.00 \footnotesize{$\pm$ 0.0} & 0.00 \footnotesize{$\pm$ 0.0} & 0.00 \footnotesize{$\pm$ 0.0} & 0.00 \footnotesize{$\pm$ 0.0} & 0.00 \footnotesize{$\pm$ 0.0} & 0.00 \footnotesize{$\pm$ 0.0} \\
CFKG & \underline{5.00} \footnotesize{$\pm$ 0.2} & 0.20 \footnotesize{$\pm$ 0.0} & \underline{9.96} \footnotesize{$\pm$ 0.9} & 1.23 \footnotesize{$\pm$ 0.1} & 0.00 \footnotesize{$\pm$ 0.0} & 0.00 \footnotesize{$\pm$ 0.0} & 0.00 \footnotesize{$\pm$ 0.0} & 0.00 \footnotesize{$\pm$ 0.0} & 0.27 \footnotesize{$\pm$ 0.2} & 0.18 \footnotesize{$\pm$ 0.1} \\
KGCN & 0.34 \footnotesize{$\pm$ 0.3} & 0.01 \footnotesize{$\pm$ 0.0} & 0.15 \footnotesize{$\pm$ 0.1} & 0.00 \footnotesize{$\pm$ 0.0}  & \underline{1.02} \footnotesize{$\pm$ 1.3} & 0.02 \footnotesize{$\pm$ 0.0} & 0.20 \footnotesize{$\pm$ 0.1} & 0.01 \footnotesize{$\pm$ 0.0} & \underline{0.40} \footnotesize{$\pm$ 0.3} & 0.01 \footnotesize{$\pm$ 0.0} \\
MKR$^\text{\color{NavyBlue}\tiny\SnowflakeChevron}$ & 0.00 \footnotesize{$\pm$ 0.0} & 0.00 \footnotesize{$\pm$ 0.0} & 0.00 \footnotesize{$\pm$ 0.0} & 0.00 \footnotesize{$\pm$ 0.0}  & 0.00 \footnotesize{$\pm$ 0.0} & 0.00 \footnotesize{$\pm$ 0.0} & 0.00 \footnotesize{$\pm$ 0.0} & 0.00 \footnotesize{$\pm$ 0.0} & 0.00 \footnotesize{$\pm$ 0.0} & 0.00 \footnotesize{$\pm$ 0.0} \\
SpectralCF$^\text{\color{NavyBlue}\tiny\SnowflakeChevron}$ & 0.00 \footnotesize{$\pm$ 0.0} & 0.00 \footnotesize{$\pm$ 0.0} & 0.00 \footnotesize{$\pm$ 0.0} & 0.00 \footnotesize{$\pm$ 0.0} & 0.00 \footnotesize{$\pm$ 0.0} & 0.00 \footnotesize{$\pm$ 0.0} & 0.00 \footnotesize{$\pm$ 0.0} & 0.00 \footnotesize{$\pm$ 0.0} & 0.00 \footnotesize{$\pm$ 0.0} & 0.00 \footnotesize{$\pm$ 0.0} \\
\midrule
PGPR$_a$ & \textbf{46.03} \footnotesize{$\pm$ 1.2} & 10.27 \footnotesize{$\pm$ 0.3} & \textbf{28.98} \footnotesize{$\pm$ 0.9} & 7.35 \footnotesize{$\pm$ 0.4} & \textbf{39.79} \footnotesize{$\pm$ 1.3} & 11.18 \footnotesize{$\pm$ 0.3} & \textbf{40.66} \footnotesize{$\pm$ 1.7} & 9.59 \footnotesize{$\pm$ 0.3} & 21.88 \footnotesize{$\pm$ 1.8} & 0.88 \footnotesize{$\pm$ 0.0} \\
PGPR$_0$ & 2.85 \footnotesize{$\pm$ 0.2} & 0.05 \footnotesize{$\pm$ 0.0} & 0.08 \footnotesize{$\pm$ 0.1} & 0.00 \footnotesize{$\pm$ 0.0} & 1.58 \footnotesize{$\pm$ 0.1} & 0.02 \footnotesize{$\pm$ 0.0} & 0.97 \footnotesize{$\pm$ 0.3} & 0.03 \footnotesize{$\pm$ 0.0} & 10.86 \footnotesize{$\pm$ 0.7} & 0.25 \footnotesize{$\pm$ 0.0} \\
UPGPR$_a$ & 33.10 \footnotesize{$\pm$ 3.1} & \textbf{11.46} \footnotesize{$\pm$ 1.2} & 23.96 \footnotesize{$\pm$ 3.0} & 5.70 \footnotesize{$\pm$ 0.5} & 35.83 $\pm$ 4.2 & \textbf{11.97} \footnotesize{$\pm$ 2.2} & 31.72 $\pm$ 5.2 & 8.34 \footnotesize{$\pm$ 0.4} & \textbf{27.72} \footnotesize{$\pm$ 2.8} & 3.52 \footnotesize{$\pm$ 0.2} \\
UPGPR$_0$ & 21.05 \footnotesize{$\pm$ 1.1} & 2.96 \footnotesize{$\pm$ 0.1} & 2.31 \footnotesize{$\pm$ 0.0} & 0.34 \footnotesize{$\pm$ 0.0} & 14.41 $\pm$ 1.6 & 2.07 \footnotesize{$\pm$ 0.0} & 11.09 \footnotesize{$\pm$ 1.0} & 1.47 \footnotesize{$\pm$ 0.1} & 19.15 \footnotesize{$\pm$ 0.6} & 2.86 \footnotesize{$\pm$ 0.3} \\
\bottomrule
\end{tabular}
}
\caption{Perfomance of \grecs compared to the baselines for strict cold start items on all datasets. The best results are highlighted in bold and the best baseline is underlined.}
\label{tab:all_cold_item_results}
\vspace{-6mm}
\end{table*}

Table~\ref{tab:all_cold_item_results} details the performance of \grecs compared to baselines for strictly cold items. We observe that: \textbf{(1)} \grecs outperforms all baselines in cold item coverage across all five datasets, indicating that \grecs is capable of identifying a wider variety of relevant cold items across different users than baseline approaches.
\textbf{(2)} ItemKNN shows high Prop.$^\text{\color{NavyBlue}\tiny\SnowflakeChevron}$, matching or exceeding \grecs on four out of the five datasets. ItemKNN recommends on average one cold item in the top 10 for each user. However, the low coverage of ItemKNN (less than $1\%$ on all datasets) indicates that the method tends to recommend the same cold items to all users, revealing a lack of personalization.
\textbf{(3)} UPGPR$_a$/PGPR$_a$ significantly outperform UPGPR$_0$/PGPR$_0$ in terms of coverage and proportion, suggesting that agents avoid recommending items with zero-initialized cold embeddings. Despite this observation, the null embedding strategy still manages to recommend relevant items to cold users (see Table~\ref{tab:all_cold_results}), as the agents have to start from these users to provide a recommendation (regardless of their embedding values). Combining the results from Tables~\ref{tab:all_cold_results} and~\ref{tab:all_cold_item_results}, the average translation strategy for cold embeddings is the most effective for covering both cold user and cold item scenarios.
\begin{graybox} \grecs archives the highest cold item coverage, indicating its ability to recommend cold items.\end{graybox}

\vspace{-3mm}
\subsection{Exp. 3: \grecs mitigates the popularity bias}
\label{subsec:C3}
\vspace{-1mm}

\begin{wraptable}{r}{0.5\textwidth}
\vspace{-1mm}
\centering
\resizebox{0.5\textwidth}{!}{%
\begin{tabular}{l|c|c|c|c|c}
\multicolumn{1}{c}{} & \multicolumn{1}{c}{\textbf{Beauty}} & \multicolumn{1}{c}{\textbf{CDs}} & \multicolumn{1}{c}{\textbf{Cellphones}} & \multicolumn{1}{c}{\textbf{Clothing}} & \multicolumn{1}{c}{\textbf{COCO}}\\
\toprule
Pop & 1.00 \footnotesize{$\pm$ 0.0} & 1.00 \footnotesize{$\pm$ 0.0} & 1.00 \footnotesize{$\pm$ 0.0} & 1.00 \footnotesize{$\pm$ 0.0} & 1.00 \footnotesize{$\pm$ 0.0} \\
ItemKNN & \underline{\textbf{0.06}} \footnotesize{$\pm$ 0.0} & \underline{\textbf{0.05}} \footnotesize{$\pm$ 0.0} & \underline{\textbf{0.03}} \footnotesize{$\pm$ 0.0} & \underline{\textbf{0.02}} \footnotesize{$\pm$ 0.0} & \underline{0.13} \footnotesize{$\pm$ 0.0} \\
BPR & 0.32 \footnotesize{$\pm$ 0.0} & 0.21 \footnotesize{$\pm$ 0.0} & 0.33 \footnotesize{$\pm$ 0.0} & 0.26 \footnotesize{$\pm$ 0.0} & 0.23 \footnotesize{$\pm$ 0.0} \\
NeuMF & 0.36 \footnotesize{$\pm$ 0.0} & 0.21 \footnotesize{$\pm$ 0.0} & 0.37 \footnotesize{$\pm$ 0.0} & 0.39 \footnotesize{$\pm$ 0.0} & 0.25 \footnotesize{$\pm$ 0.0} \\
CFKG & 0.39 \footnotesize{$\pm$ 0.0} & 0.28 \footnotesize{$\pm$ 0.0} & 0.26 \footnotesize{$\pm$ 0.0} & 0.22 \footnotesize{$\pm$ 0.0} & 0.47 \footnotesize{$\pm$ 0.1} \\
KGCN & 0.34 \footnotesize{$\pm$ 0.0} & 0.16 \footnotesize{$\pm$ 0.0} & 0.38 \footnotesize{$\pm$ 0.1} & 0.31 \footnotesize{$\pm$ 0.0} & 0.25 \footnotesize{$\pm$ 0.0} \\
MKR$^\text{\color{NavyBlue}\tiny\SnowflakeChevron}$ & 0.42 \footnotesize{$\pm$ 0.1} &  0.31 \footnotesize{$\pm$ 0.1}  & 0.52 \footnotesize{$\pm$ 0.0} & 0.43 \footnotesize{$\pm$ 0.1} & 0.36 \footnotesize{$\pm$ 0.1} \\
SpectralCF$^\text{\color{NavyBlue}\tiny\SnowflakeChevron}$ & 0.75 \footnotesize{$\pm$ 0.2} & 0.25 \footnotesize{$\pm$ 0.0} & 0.96 \footnotesize{$\pm$ 0.0} & 0.83 \footnotesize{$\pm$ 0.1} & 0.90 \footnotesize{$\pm$ 0.0} \\
\midrule
PGPR$_a$ & 0.27 \footnotesize{$\pm$ 0.0} & 0.37 \footnotesize{$\pm$ 0.0} & 0.36 \footnotesize{$\pm$ 0.0} & 0.27 \footnotesize{$\pm$ 0.0} & 0.24 \footnotesize{$\pm$ 0.0} \\
PGPR$_0$ & 0.32 \footnotesize{$\pm$ 0.0} & 0.41 \footnotesize{$\pm$ 0.0} & 0.51 \footnotesize{$\pm$ 0.0} & 0.46 \footnotesize{$\pm$ 0.0} & 0.27 \footnotesize{$\pm$ 0.0} \\
UPGPR$_a$ & 0.12 \footnotesize{$\pm$ 0.0} & 0.10 \footnotesize{$\pm$ 0.0} & 0.09 \footnotesize{$\pm$ 0.0} & 0.12 \footnotesize{$\pm$ 0.0} & \textbf{0.05} \footnotesize{$\pm$ 0.0} \\
UPGPR$_0$ & 0.15 \footnotesize{$\pm$ 0.0} & 0.11 \footnotesize{$\pm$ 0.0} & 0.11 \footnotesize{$\pm$ 0.0} & 0.14 \footnotesize{$\pm$ 0.0} & \textbf{0.05} \footnotesize{$\pm$ 0.0} \\
\bottomrule
\end{tabular}
}
\caption{Popularity Bias (POPB) of \grecs compared to the baselines on all datasets. The best results (lowest) are highlighted in bold and the best baseline is underlined.}
\label{tab:pop-bias}
\vspace{-3mm}
\end{wraptable}

Table~\ref{tab:pop-bias} shows the popularity bias of \grecs compared to the baselines across all datasets. \
Interestingly, ItemKNN exhibits the lowest bias because it relies only on items each user has interacted with, it is less sensitive to the popularity bias than supervised methods optimized on the entire dataset. 
Despite this low bias, ItemKNN performs poorly in scenarios with cold users (as shown in Table~\ref{tab:all_cold_results}).
UPGPR is the second-best method, consistently outperforming all other baselines as well as PGPR.
These results suggest that \grecs is better suited to handle popularity bias than the baseline approaches, due to its exploration of the KG, which restricts the set of recommendable items.
The large difference between PGPR and UPGPR indicates that the reward mechanism greatly impacts mitigating popularity bias in RL-based GR methods. PGPR relies on embedding-similarity-based rewards, while UPGPR relies solely on the relevance of the recommendation, leading to less biased outcomes.
Finally, PGPR$_0$/UPGPR$_0$ are more bias towards popular items than PGPR$_a$/UPGPR$_a$. 
This finding highlights the importance of embedding initialization in mitigating the popularity bias, showing that using neighboring embeddings instead of null embeddings makes recommendations less biased.

\begin{graybox} \grecs archives the best trade-off between unbiased and relevant recommendations.\end{graybox}

\vspace{-1mm}
\subsection{Exp. 4: \grecs requires few relations to provide accurate recommendations}
\label{subsec:C4}
Lastly, we evaluated the effectiveness of \grecs for sparse cold start scenarios with few interactions and explored the impact of the number of known relations on performance for strict cold start users.
\vspace{-1mm}
\begin{figure}[t]
  \begin{subfigure}{.2\textwidth}
  \centering
    \includegraphics[width=1\linewidth,height=2.75cm]{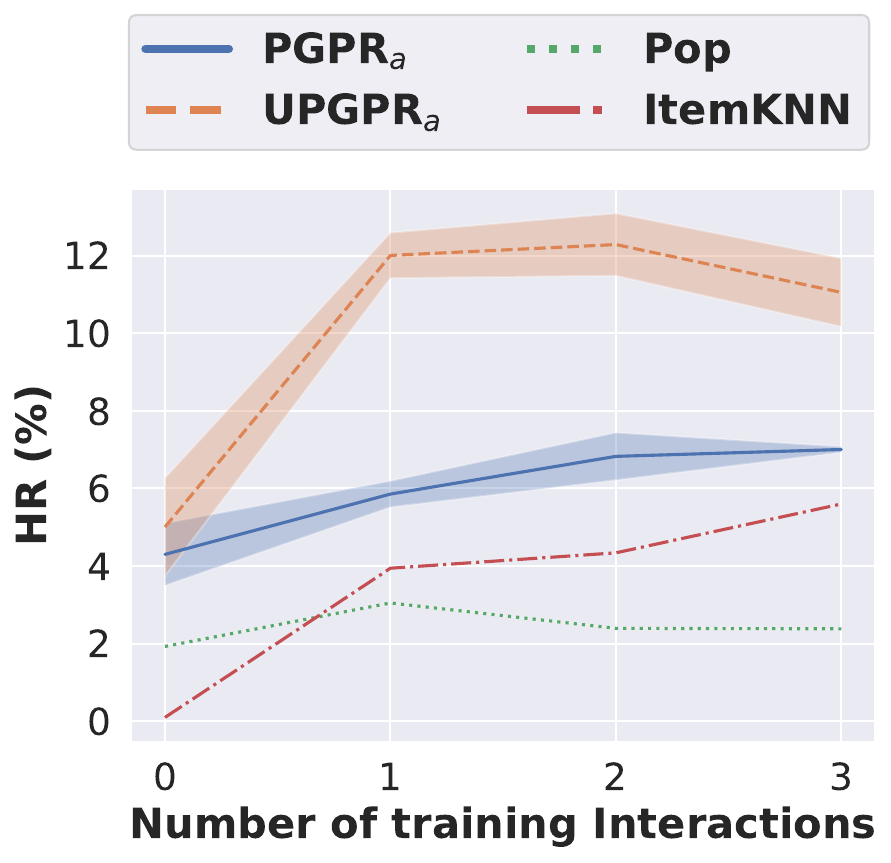}
    \caption{Beauty}
  \end{subfigure}%
  \begin{subfigure}{.2\textwidth}
  \centering
    \includegraphics[width=1\linewidth,height=2.75cm]{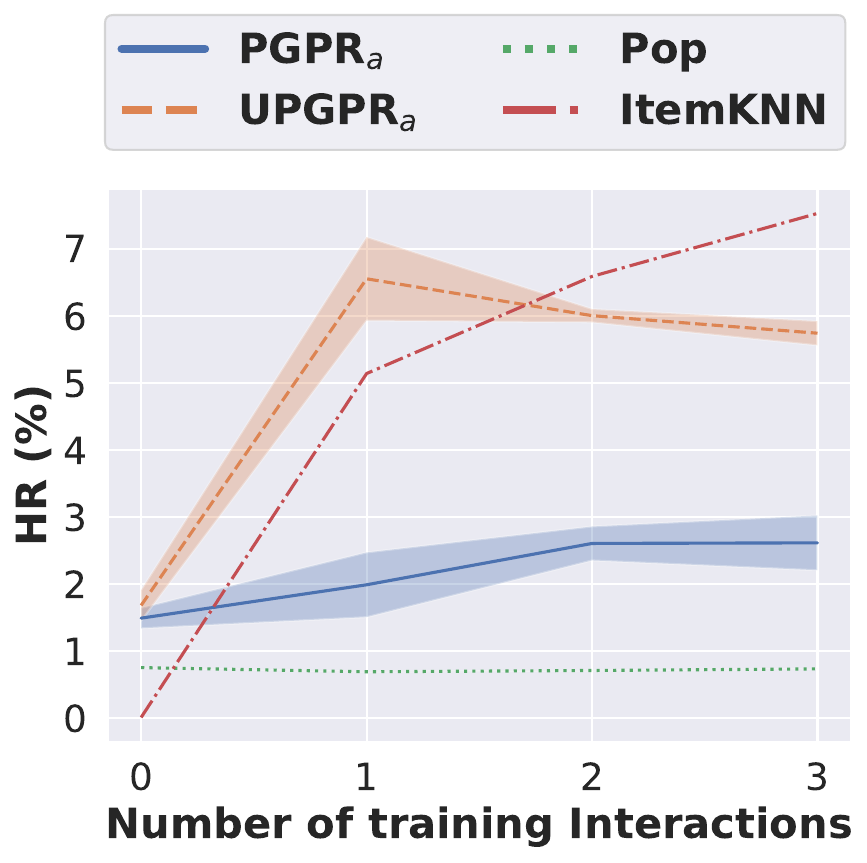}
    \caption{CDs}
  \end{subfigure}%
  \begin{subfigure}{.2\textwidth}
  \centering
    \includegraphics[width=1\linewidth,height=2.75cm]{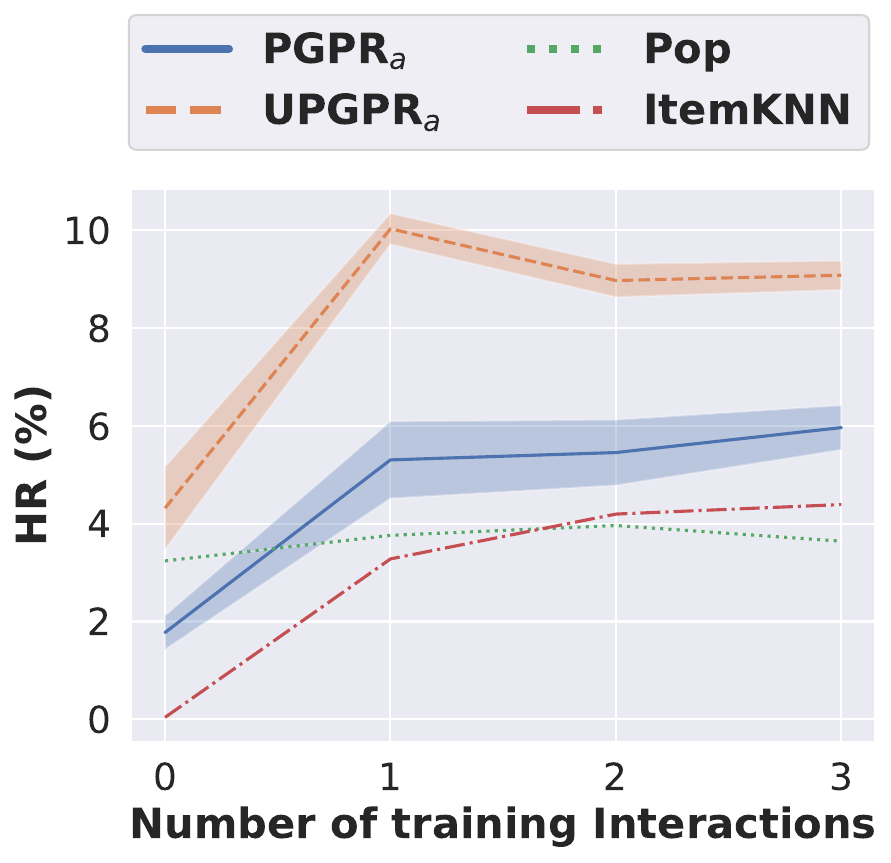}
    \caption{Cellphones}
  \end{subfigure}%
  \begin{subfigure}{.2\textwidth}
  \centering
    \includegraphics[width=1\linewidth,height=2.75cm]{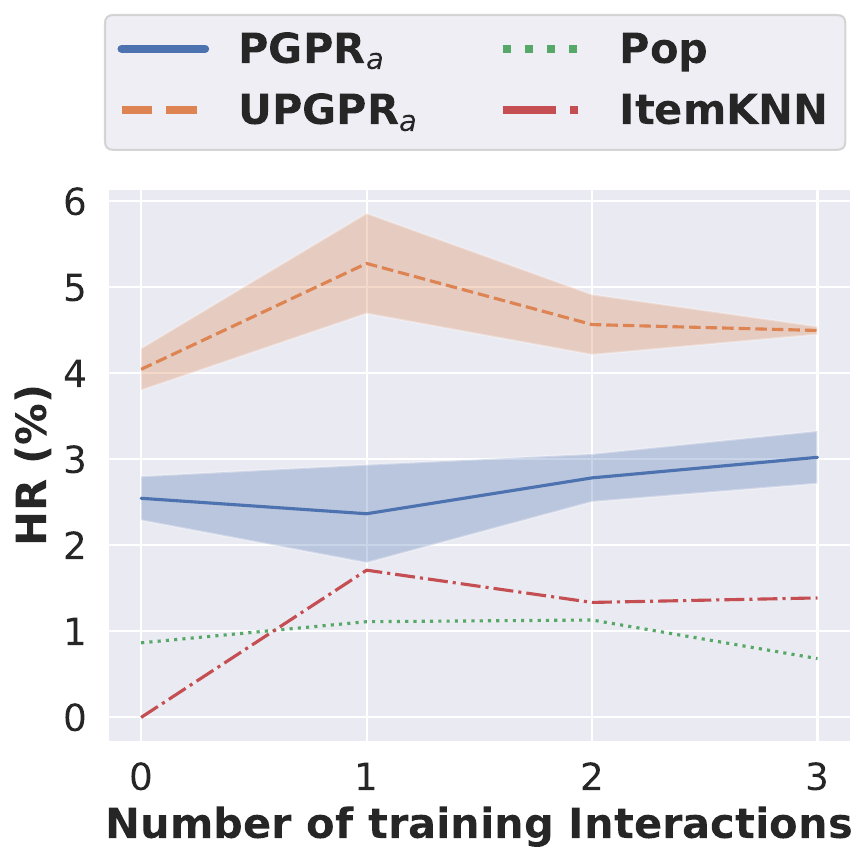}
    \caption{Clothing}
  \end{subfigure}%
  \begin{subfigure}{.2\textwidth}
  \centering
    \includegraphics[width=1\linewidth,height=2.75cm]{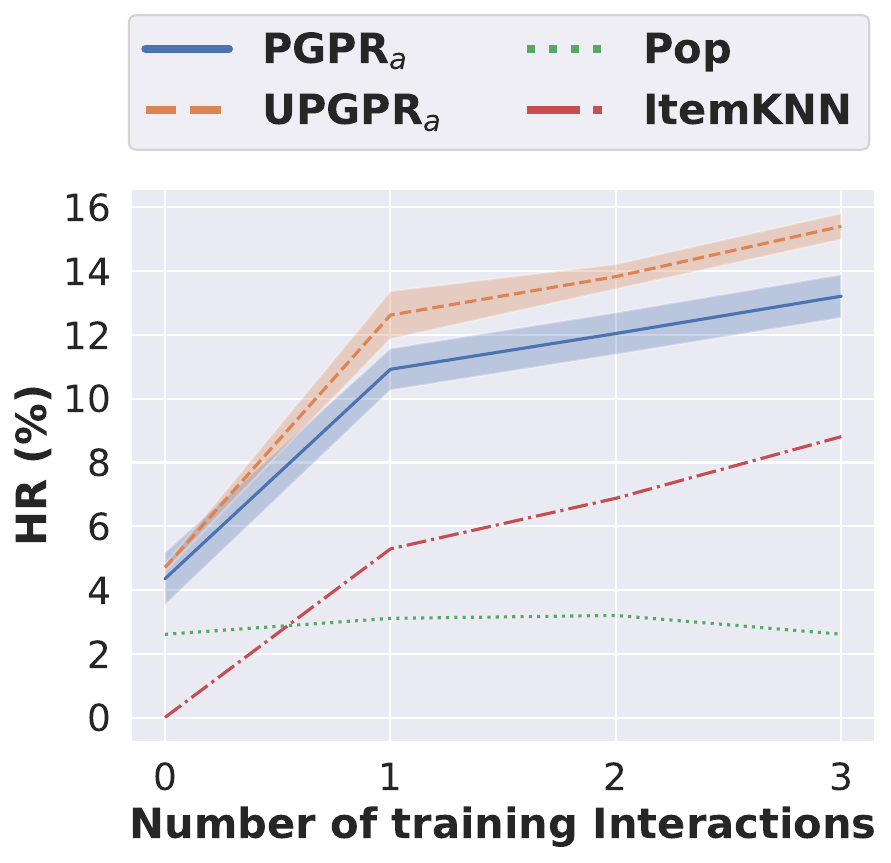}
    \caption{COCO}
  \end{subfigure}%
  \caption{Evolution of the HR against the number of training interactions. The shaded area represents the standard deviation across 3 runs.}
  \label{fig:nb_train}
  \vspace{-3mm}
\end{figure}

 Figure~\ref{fig:nb_train} shows the HR against the number of interactions in the training set, ranging from 0 (strict cold start) to 3. We only display the Pop and BPR baselines, as they perform the best for strict cold start and warm users respectively. As expected, the performance of the Pop baseline does not improve with the number of interactions, since it recommends the most popular items regardless of the user's profile. Conversely, all other approaches show improvement, even from minimal interaction data. The results show that even a single known interaction in the training set can significantly enhance model performance. Fianlly, we see that UPGPR$_a$ surpasses PGPR$_a$ in performance on the E-commerce datasets, while both methods exhibit comparable results on the COCO dataset.

\begin{figure}[t]
  \begin{subfigure}{.2\textwidth}
  \centering
    \includegraphics[width=1\linewidth,height=2.5cm]{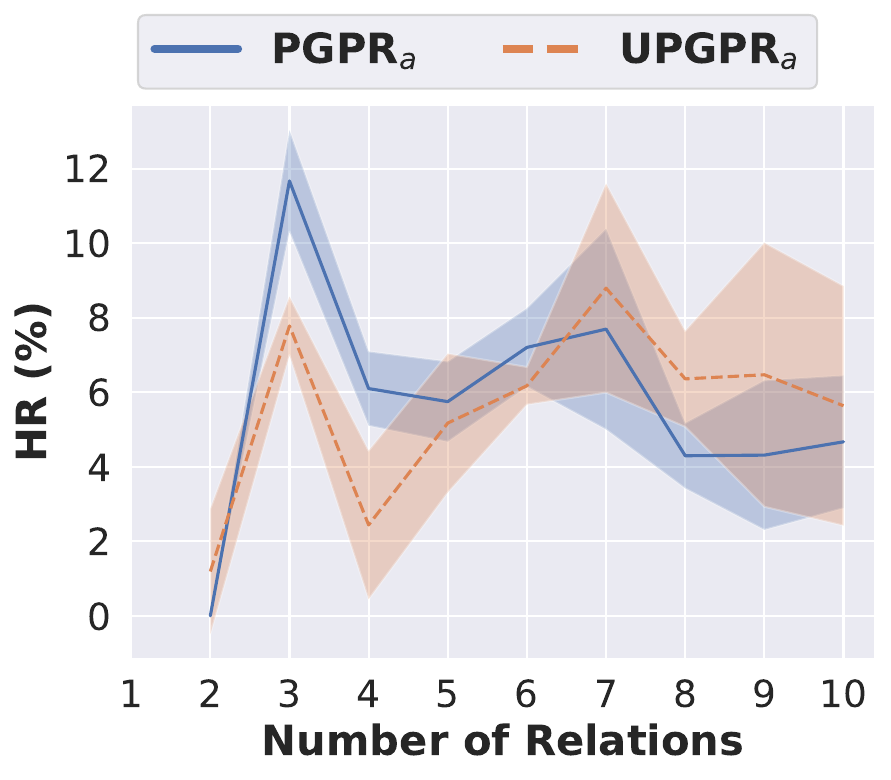}
    \caption{Beauty}
  \end{subfigure}%
  \begin{subfigure}{.2\textwidth}
  \centering
    \includegraphics[width=1\linewidth,height=2.5cm]{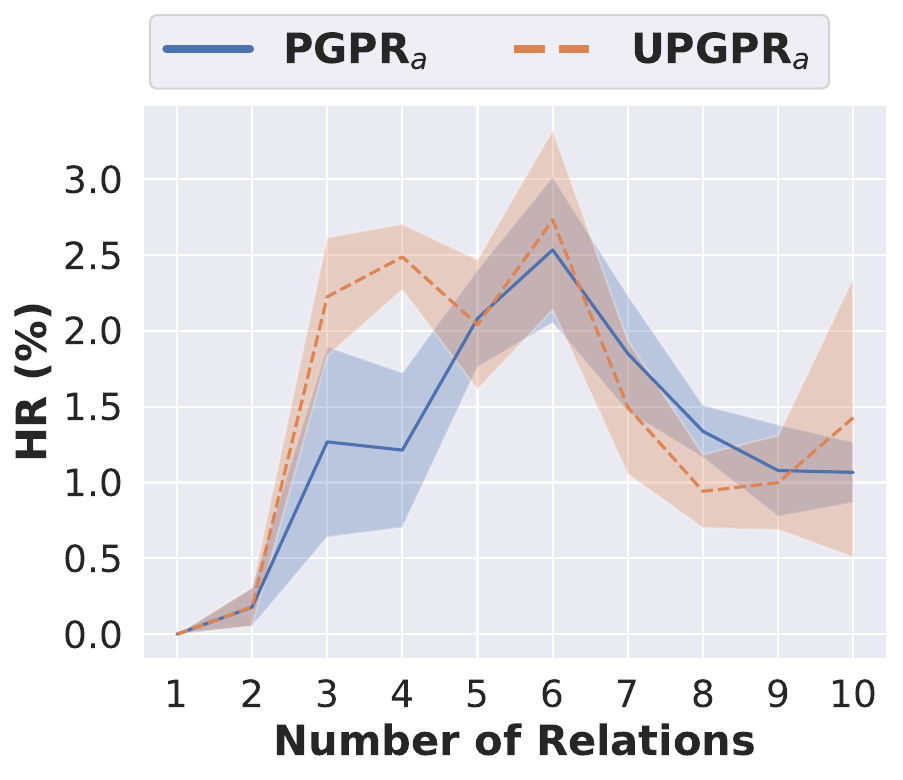}
    \caption{CDs}
  \end{subfigure}%
  \begin{subfigure}{.2\textwidth}
  \centering
    \includegraphics[width=1\linewidth,height=2.5cm]{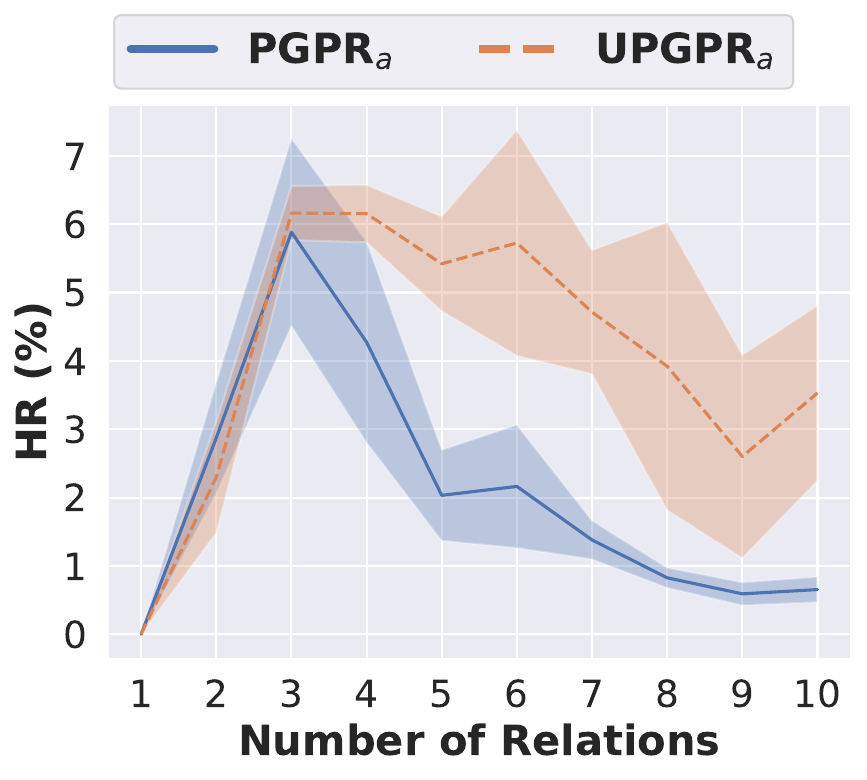}
    \caption{Cellphones}
  \end{subfigure}%
  \begin{subfigure}{.2\textwidth}
  \centering
    \includegraphics[width=1\linewidth,height=2.5cm]{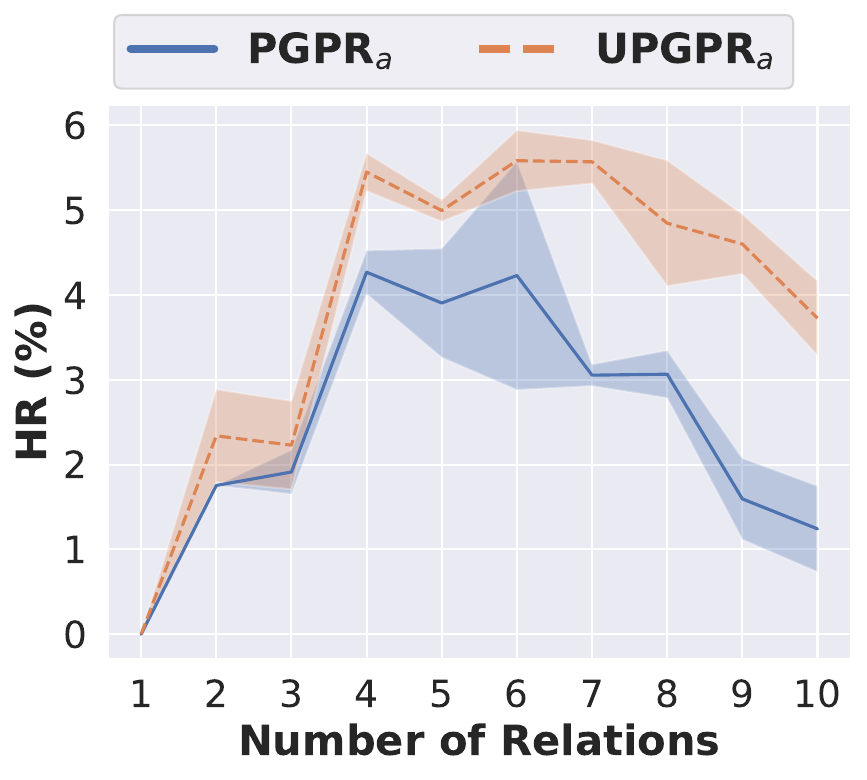}
    \caption{Clothing}
  \end{subfigure}%
  \begin{subfigure}{.2\textwidth}
  \centering
    \includegraphics[width=1\linewidth,height=2.5cm]{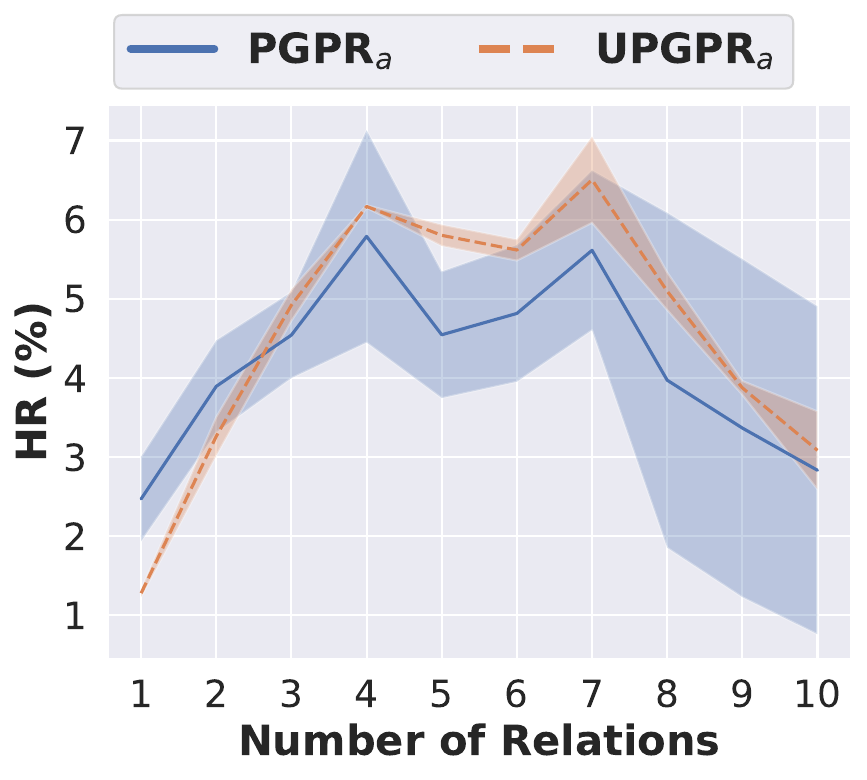}
    \caption{COCO}
  \end{subfigure}%
  \caption{Evolution of the HR against the number of relations for strict cold start users. The shaded area represents the standard deviation across 3 runs.}
  \label{fig:nb_rel}
  \vspace{-3mm}
\end{figure}

Figure~\ref{fig:nb_rel} shows the HR relative to the number of known relations for strict cold start users. For instance, in an E-Commerce dataset, a strict cold start user without any \textit{purchase} interactions still has relations such as \textit{like} or \textit{interested\_in}, indicating preferences for categories or brands. Generally, \grecs benefits from an increase in known relations for strict cold start users, with HRs improving initially with the addition of a few relations. however, HR declines and its variance increases with too many known relations and no interactions. This is expected: recommending items to a cold user who likes a variety of brands is more difficult than recommending items to a user with specific preferences.

\begin{graybox} \grecs is best suited for strict cold start recommendations when a small number of relations are known, and knowing a few interactions increases the relevance of recommendations even further.\end{graybox}


\vspace{-3mm}
\section{Discussion \& Conclusion}
\label{sec:conclusion}
\vspace{-3mm}
In this paper, we proposed \grecs, a framework for adapting GR-based recommendations to cold start scenarios while keeping recommendations explainable. By explicitly using relationships in the KG to connect users to items, \grecs can find items matching users’ preferences even with limited information about users. Our experiments across five datasets show that \grecs provides to better and less biased recommendations for new users than state-of-the-art baselines. \grecs is also able to recommend cold items, requiring only few relations to make relevant recommendations. Furthermore, we show that both the strategy for initializing the embeddings of cold-start users and items as well as the choice of reward are crucial for recommending cold items and mitigating the popularity bias.

\textbf{Limitations and future work.} \grecs relies on KGs and can therefore make recommendations without pre-existing relational data about cold users or items. It can for example not take advantage of attributes with real values such as the item's price. Furthermore, the GR methods at the basis of \grecs exhibit an inherent recall bias~\cite{geng2022path}. 
Specifically, their recall is limited by the existence of user-item pairs that cannot be connected through paths of length $k$, restricting their ability to recommend certain relevant items.
Finally, our cold embedding assignment method is specific to PGPR and UPGPR. Future work should extend \grecs to other GR approaches, namely LLM-based approaches that are not subject to the recall bias.

\bibliographystyle{ACM-Reference-Format}
\bibliography{references}


\newpage
\section*{NeurIPS Paper Checklist}

\begin{enumerate}

\item {\bf Claims}
    \item[] Question: Do the main claims made in the abstract and introduction accurately reflect the paper's contributions and scope?
    \item[] Answer: \answerYes{} 
    \item[] Justification: We claim that Graph Reasoning methods are effective at addressing the cold start problem. Our experiments show that this is the case.

\item {\bf Limitations}
    \item[] Question: Does the paper discuss the limitations of the work performed by the authors?
    \item[] Answer: \answerYes{} 
    \item[] Justification: We discuss the limitations of our work in Section~\ref{sec:conclusion}.

\item {\bf Theory Assumptions and Proofs}
    \item[] Question: For each theoretical result, does the paper provide the full set of assumptions and a complete (and correct) proof?
    \item[] Answer: \answerNA{} 
    \item[] Justification: We do not have theoretical results in this work.

    \item {\bf Experimental Result Reproducibility}
    \item[] Question: Does the paper fully disclose all the information needed to reproduce the main experimental results of the paper to the extent that it affects the main claims and/or conclusions of the paper (regardless of whether the code and data are provided or not)?
    \item[] Answer: \answerYes{} 
    \item[] Justification: We provided the libraries and hyperparameter values used in our experiments and indicated when we used default values and where to find them. We also indicated the pre-processing and splitting strategies performed on our datasets. 

\item {\bf Open access to data and code}
    \item[] Question: Does the paper provide open access to the data and code, with sufficient instructions to faithfully reproduce the main experimental results, as described in supplemental material?
    \item[] Answer: \answerYes{} 
    \item[] Justification: Our code is publicly available as well as the datasets we performed our experiments on.

\item {\bf Experimental Setting/Details}
    \item[] Question: Does the paper specify all the training and test details (e.g., data splits, hyperparameters, how they were chosen, type of optimizer, etc.) necessary to understand the results?
    \item[] Answer: \answerYes{} 
    \item[] Justification: We discuss the experiential setting in Section~\ref{sec:experiments}.

\item {\bf Experiment Statistical Significance}
    \item[] Question: Does the paper report error bars suitably and correctly defined or other appropriate information about the statistical significance of the experiments?
    \item[] Answer: \answerYes{} 
    \item[] Justification: We ran experiments across multiple seeds and reported the standard deviation on tables as a number and on graphs as a shaded area.

\item {\bf Experiments Compute Resources}
    \item[] Question: For each experiment, does the paper provide sufficient information on the computer resources (type of compute workers, memory, time of execution) needed to reproduce the experiments?
    \item[] Answer: \answerYes{} 
    \item[] Justification: We specified the CPU model and time of execution in Section~\ref{sec:results}
    
\item {\bf Code Of Ethics}
    \item[] Question: Does the research conducted in the paper conform, in every respect, with the NeurIPS Code of Ethics \url{https://neurips.cc/public/EthicsGuidelines}?
    \item[] Answer: \answerYes{} 
    \item[] Justification: Our experiments did not involve human participants and were performed on public, anonymized datasets.

\item {\bf Broader Impacts}
    \item[] Question: Does the paper discuss both potential positive societal impacts and negative societal impacts of the work performed?
    \item[] Answer: \answerYes{} 
    \item[] Justification: Because the models we study are explainable, we consider that their potential societal impact is positive. In the introduction, we mention the importance of providing explanations for recommendations in high-impact domains. 
    
\item {\bf Safeguards}
    \item[] Question: Does the paper describe safeguards that have been put in place for responsible release of data or models that have a high risk for misuse (e.g., pretrained language models, image generators, or scraped datasets)?
    \item[] Answer: \answerNA{} 
    \item[] Justification: Our work presents no such risk, the models can only make recommendations on the datasets they have been trained on, which are anonymized.

\item {\bf Licenses for existing assets}
    \item[] Question: Are the creators or original owners of assets (e.g., code, data, models), used in the paper, properly credited and are the license and terms of use explicitly mentioned and properly respected?
    \item[] Answer: \answerYes{} 
    \item[] Justification: We cite the relevant papers and provide the URLs for both the code and datasets.

\item {\bf New Assets}
    \item[] Question: Are new assets introduced in the paper well documented and is the documentation provided alongside the assets?
    \item[] Answer: \answerYes{}
    \item[] Justification: The code is commented and the overall project is described in the README.md in \url{https://anonymous.4open.science/r/cold_rec-B765}.

\item {\bf Crowdsourcing and Research with Human Subjects}
    \item[] Question: For crowdsourcing experiments and research with human subjects, does the paper include the full text of instructions given to participants and screenshots, if applicable, as well as details about compensation (if any)? 
    \item[] Answer: \answerNA{} 
    \item[] Justification: The paper does not involve crowdsourcing nor research with human subjects.

\item {\bf Institutional Review Board (IRB) Approvals or Equivalent for Research with Human Subjects}
    \item[] Question: Does the paper describe potential risks incurred by study participants, whether such risks were disclosed to the subjects, and whether Institutional Review Board (IRB) approvals (or an equivalent approval/review based on the requirements of your country or institution) were obtained?
    \item[] Answer: \answerNA{} 
    \item[] Justification: The paper does not involve crowdsourcing nor research with human subjects.

\end{enumerate}

\newpage

\appendix


\section{Datasets Statistics}
\label{sec:appendix_datasets}

\begin{table}[h]
\centering
\resizebox{\textwidth}{!}{%

\begin{tabular}{lcccc|lc}
\toprule
\multicolumn{5}{c}{\large\textbf{E-commerce}} & \multicolumn{2}{c}{\large\textbf{Education}}\\
\midrule
 & \textbf{CDs} & \textbf{Cloth.} & \textbf{Cell.} & \textbf{Beauty} & \multicolumn{2}{c}{\textbf{COCO}}\\ 
\midrule
\textbf{\#Entities} &       &       &       &  & \textbf{\#Entities} & \\ 
User             & 75k   & 39k   & 27k   & 22k & User  & 25k \\ 
Product  & 64k  & 23k & 10k & 12k & Course  & 23k \\  
Word   & 202k & 21k & 22k & 22k & Category   & 132  \\  
Brand  & 1.4k  & 1.1k & 955 & 2k &  Teacher  & 4.1k    \\ 
Category  & 770  & 1.1k & 206 & 248 &  Skill  & 5.5k  \\  
\midrule
\textbf{\#Relations} &  &  & &  & \textbf{\#Relations} & \\ 
User $\xrightarrow{\text{purchase}}$ Product & 1.1M  & 278k & 194k & 198k & User $\xrightarrow{\text{enroll}}$ Course & 428k\\ 
User $\xrightarrow{\text{mention}}$ Word & 191M  & 17M & 18M & 18M & User $\xrightarrow{\text{know}^\text{\color{NavyBlue}\tiny\SnowflakeChevron}}$ Skill & 278k \\
User $\xrightarrow{\text{like}^\text{\color{NavyBlue}\tiny\SnowflakeChevron}}$ Brand & 192k & 60k & 90k & 132k & Course $\xrightarrow{\text{belong\_to}^\text{\color{NavyBlue}\tiny\SnowflakeChevron}}$ Category & 23k \\ 
User $\xrightarrow{\text{interested\_in}^\text{\color{NavyBlue}\tiny\SnowflakeChevron}}$ Category & 2.0M  & 949k & 288k & 354k & Course $\xrightarrow{\text{cover}^\text{\color{NavyBlue}\tiny\SnowflakeChevron}}$ Skill & 47k \\ 
Product $\xrightarrow{\text{described\_by}}$ Word & 191M & 17M & 18M & 18M &   Course $\xrightarrow{\text{taught\_by}^\text{\color{NavyBlue}\tiny\SnowflakeChevron}}$ Teacher & 23k  \\
Product $\xrightarrow{\text{belong\_to}^\text{\color{NavyBlue}\tiny\SnowflakeChevron}}$ Category & 466k & 154k & 36k & 49k &  &  \\ 
Product $\xrightarrow{\text{produced\_by}^\text{\color{NavyBlue}\tiny\SnowflakeChevron}}$ Brand & 64k & 23k & 10k & 12k &  &  \\ 
Product $\xrightarrow{\text{also\_bought}}$ Product & 3.6M & 1.4M & 590k & 891k &  &  \\
Product $\xrightarrow{\text{also\_viewed}}$ Product & 78k & 147k & 22k & 155k &  &  \\
Product $\xrightarrow{\text{bought\_together}}$ Product & 78k & 28k & 12k & 14k & & \\
\bottomrule
\end{tabular}
}
\caption{Statistics of the e-commerce and education datasets. Relations indicated with $^\text{\color{NavyBlue}\tiny\SnowflakeChevron}$ are used to integrate cold users or cold items into the KG.}
\label{table:dataset_statistics}
\vspace{-3mm}
\end{table}

\section{Results on all users}
\label{sec:appendix_all}

Graph Reasoning approaches match or outperform all the baselines on 4 out of the 5 datasets for warm users whose interactions are available in the training set. 
These results confirm the findings of \citet{xian2019reinforcement} and  \citet{frej2024finding}: Graph Reasoning methods can make explainable and relevant recommendations competitively even against non-interpretable baselines. 
It should be noted that the results reported here are different than the ones reported in the PGPR~\cite{xian2019reinforcement} and UPGPR~\cite{frej2024finding} papers because our method for splitting the datasets is different than theirs. 
We adopted a different splitting method (user level splitting, and removing users and items from the training set) to study the recommendations under cold start scenarios.

\begin{table*}[h]
\centering
\resizebox{\textwidth}{!}{%
\begin{tabular}{l|cc|cc|cc|cc|cc}
\multicolumn{1}{c}{} & \multicolumn{2}{c}{\textbf{Beauty}} & \multicolumn{2}{c}{\textbf{CDs}} & \multicolumn{2}{c}{\textbf{Cellphones}} & \multicolumn{2}{c}{\textbf{Clothing}} & \multicolumn{2}{c}{\textbf{COCO}}\\
Measures (\%) & nDCG & HR & nDCG & HR & nDCG & HR & nDCG & HR & nDCG & HR\\
\hline
Pop & 1.00 $\pm$ 0.0 & 1.97 $\pm$ 0.0 & 0.29 $\pm$ 0.0 & 0.71 $\pm$ 0.0 & 1.60 $\pm$ 0.0 & 3.18 $\pm$ 0.0 & 0.40 $\pm$ 0.0 & 0.78 $\pm$ 0.0 & 0.98 $\pm$ 0.0 & 2.30 $\pm$ 0.0 \\
ItemKNN & 3.02 $\pm$ 0.0 & 5.58 $\pm$ 0.0 & \underline{\textbf{3.15}} $\pm$ 0.0 & \underline{\textbf{6.28}} $\pm$ 0.0 & 2.19 $\pm$ 0.0 & 4.10 $\pm$ 0.0 & 0.75 $\pm$ 0.0 & 1.33 $\pm$ 0.0 & \underline{2.90} $\pm$ 0.0 & \underline{6.03} $\pm$ 0.0 \\
BPR & \underline{3.13} $\pm$ 0.1 & \underline{6.02} $\pm$ 0.2 & 2.45 $\pm$ 0.0 & 5.37 $\pm$ 0.0 & \underline{3.32} $\pm$ 0.0 & \underline{6.10} $\pm$ 0.1 & \underline{0.84} $\pm$ 0.0 & \underline{1.54} $\pm$ 0.0 & 2.54 $\pm$ 0.0 & 5.37 $\pm$ 0.1 \\
NeuMF & 2.28 $\pm$ 0.0 & 4.55 $\pm$ 0.1 & 1.67 $\pm$ 0.0 & 3.75 $\pm$ 0.0 & 2.24 $\pm$ 0.1 & 4.18 $\pm$ 0.1 & 0.35 $\pm$ 0.0 & 0.68 $\pm$ 0.0 & 0.93 $\pm$ 0.1 & 2.17 $\pm$ 0.0 \\
CFKG & 2.54 $\pm$ 0.1 & 5.08 $\pm$ 0.2 & 1.88 $\pm$ 0.0 & 4.33 $\pm$ 0.0 & 1.35 $\pm$ 0.1 & 2.68 $\pm$ 0.1 & 0.27 $\pm$ 0.0 & 0.53 $\pm$ 0.0 & 1.31 $\pm$ 0.2 & 2.87 $\pm$ 0.5 \\
KGCN & 1.90 $\pm$ 0.1 & 3.93 $\pm$ 0.1 & 1.40 $\pm$ 0.0 & 3.36 $\pm$ 0.0 & 2.07 $\pm$ 0.2 & 3.94 $\pm$ 0.4 & 0.35 $\pm$ 0.0 & 0.70 $\pm$ 0.1 & 0.78 $\pm$ 0.5 & 1.81 $\pm$ 1.0 \\
MKR$^\text{\color{NavyBlue}\tiny\SnowflakeChevron}$ & 0.75 $\pm$ 0.0 & 1.58 $\pm$ 0.1 & 1.27 $\pm$ 0.0 & 2.98 $\pm$ 0.0 & 0.93 $\pm$ 0.1 & 2.02 $\pm$ 0.1 & 0.19 $\pm$ 0.0 & 0.41 $\pm$ 0.0 & 0.35 $\pm$ 0.1 & 0.89 $\pm$ 0.2 \\
SpectralCF$^\text{\color{NavyBlue}\tiny\SnowflakeChevron}$ & 0.91 $\pm$ 0.1 & 1.95 $\pm$ 0.1 & 0.55 $\pm$ 0.0 & 1.50 $\pm$ 0.0 & 1.51 $\pm$ 0.0 & 3.10 $\pm$ 0.1 & 0.35 $\pm$ 0.0 & 0.70 $\pm$ 0.1 & 0.67 $\pm$ 0.2 & 1.53 $\pm$ 0.4 \\
\hline
PGPR$_a$ & 3.08 $\pm$ 0.1 & 5.77 $\pm$ 0.0 & 1.20 $\pm$ 0.1 & 2.76 $\pm$ 0.3 & 2.53 $\pm$ 0.2 & 4.86 $\pm$ 0.4 & 1.35 $\pm$ 0.1 & 2.58 $\pm$ 0.2 & 4.21 $\pm$ 0.3 & 9.40 $\pm$ 0.6 \\
PGPR$_0$ & 3.07 $\pm$ 0.1 & 5.81 $\pm$ 0.0 & 1.17 $\pm$ 0.1 & 2.75 $\pm$ 0.3 & 1.95 $\pm$ 0.1 & 3.87 $\pm$ 0.2 & 0.65 $\pm$ 0.0 & 1.29 $\pm$ 0.0 & 4.22 $\pm$ 0.2 & 10.30 $\pm$ 0.1 \\
UPGPR$_a$ & 4.15 $\pm$ 0.2 & 8.16 $\pm$ 0.5 & 2.17 $\pm$ 0.0 & 4.26 $\pm$ 0.1 & 3.86 $\pm$ 0.1 & 7.42 $\pm$ 0.2 & \textbf{1.75} $\pm$ 0.0 & \textbf{3.85} $\pm$ 0.2 & \textbf{7.29} $\pm$ 0.2 & \textbf{12.25} $\pm$ 0.4 \\
UPGPR$_0$ & \textbf{4.30} $\pm$ 0.2 & \textbf{8.61} $\pm$ 0.3 & 2.19 $\pm$ 0.1 & 4.41 $\pm$ 0.1 & \textbf{4.07} $\pm$ 0.1 & \textbf{7.90} $\pm$ 0.3 & 1.68 $\pm$ 0.1 & 3.73 $\pm$ 0.2 & 7.26 $\pm$ 0.2 & 12.15 $\pm$ 0.3 \\
\bottomrule
\end{tabular}
}
\caption{Perfomance of Graph reasoning approaches compared to the baselines across all users on all datasets. The best results are highlighted in bold and the best results for the baselines are underlined.}
\label{tab:all_results}
\end{table*}

\section{Pattern proportion on all datasets}
\label{sec:appendix_patterns}

\begin{table*}[h]
\centering
\resizebox{\textwidth}{!}{
\begin{tabular}{lcccccccc}
 & \textbf{PGPR$_a^\text{\color{OrangeRed}\tiny\faHotjar}$} & \textbf{PGPR$_0^\text{\color{OrangeRed}\tiny\faHotjar}$} & \textbf{UPGPR$_a^\text{\color{OrangeRed}\tiny\faHotjar}$} & \textbf{UPGPR$_0^\text{\color{OrangeRed}\tiny\faHotjar}$} & \textbf{PGPR$_a^\text{\color{NavyBlue}\tiny\SnowflakeChevron}$} & \textbf{PGPR$_0^\text{\color{NavyBlue}\tiny\SnowflakeChevron}$} & \textbf{UPGPR$_a^\text{\color{NavyBlue}\tiny\SnowflakeChevron}$} & \textbf{UPGPR$_0^\text{\color{NavyBlue}\tiny\SnowflakeChevron}$} \\
 \midrule
User $\xrightarrow{\text{mentioned}}$ Word $\xrightarrow{\text{described}}$ Product & 71.58\% & 70.17\% & 46.72\% & 44.21\% & 0\% & 0\% & 0\% & 0\%\\
\rowcolor{lightcyan}
User $\xrightarrow{\text{interested\_in}^\text{\color{NavyBlue}\SnowflakeChevron}}$ Category $\xrightarrow{\text{belong\_to}^\text{\color{NavyBlue}\SnowflakeChevron}}$ Product & 16.78\% & 18.7\% & 29.18\% & 30.52\% & 65.18\% & 73.58\% & 64.95\% & 61.99\%\\
User $\xrightarrow{\text{purchase}}$ Product $\xrightarrow{\text{also\_bought}}$ Product $\xrightarrow{\text{also\_bought}}$ Product & 4.12\% & 3.38\% & 0.51\% & 0.48\% & 0\% & 0\% & 0\% & 0\%\\
User $\xrightarrow{\text{purchase}}$ Product $\xrightarrow{\text{described}}$ Word $\xrightarrow{\text{described}}$ Product & 3.98\% & 4.1\% & 0.05\% & 0.05\% & 0\% & 0\% & 0\% & 0\%\\
\rowcolor{lightcyan}
User $\xrightarrow{\text{like}^\text{\color{NavyBlue}\SnowflakeChevron}}$ Brand $\xrightarrow{\text{produced\_by}^\text{\color{NavyBlue}\SnowflakeChevron}}$ Product & 2.26\% & 2.31\% & 23.52\% & 24.73\% & 31.3\% & 25.07\% & 35.0\% & 38.01\%\\
\bottomrule
\end{tabular}
}
\caption{Proportion of the most frequent patterns used by the agents on the test set of the Beauty dataset for warm users ($^\text{\color{OrangeRed}\tiny\faHotjar}$) and cold users ($^\text{\color{NavyBlue}\tiny\SnowflakeChevron}$). Rows highlighted in blue denote cold start-compatible patterns that can be used for recommending an item to a strict cold user.}
\label{table:beauty_patterns}
\end{table*}

\begin{table*}[h]
\centering
\resizebox{\textwidth}{!}{
\begin{tabular}{lcccccccc}
 & \textbf{PGPR$_a^\text{\color{OrangeRed}\tiny\faHotjar}$} & \textbf{PGPR$_0^\text{\color{OrangeRed}\tiny\faHotjar}$} & \textbf{UPGPR$_a^\text{\color{OrangeRed}\tiny\faHotjar}$} & \textbf{UPGPR$_0^\text{\color{OrangeRed}\tiny\faHotjar}$} & \textbf{PGPR$_a^\text{\color{NavyBlue}\tiny\SnowflakeChevron}$} & \textbf{PGPR$_0^\text{\color{NavyBlue}\tiny\SnowflakeChevron}$} & \textbf{UPGPR$_a^\text{\color{NavyBlue}\tiny\SnowflakeChevron}$} & \textbf{UPGPR$_0^\text{\color{NavyBlue}\tiny\SnowflakeChevron}$} \\
 \midrule
User $\xrightarrow{\text{mentioned}}$ Word $\xrightarrow{\text{described}}$ Product & 49.09\% & 48.74\% & 63.54\% & 61.31\% & 0\% & 0\% & 0\% & 0\%\\
\rowcolor{lightcyan}
User $\xrightarrow{\text{interested\_in}^\text{\SnowflakeChevron}}$ Category $\xrightarrow{\text{belong\_to}^\text{\SnowflakeChevron}}$ Product & 30.35\% & 30.56\% & 23.24\% & 24.77\% & 85.01\% & 80.6\% & 80.3\% & 75.4\%\\
User $\xrightarrow{\text{mentioned}}$ Word $\xrightarrow{\text{mentioned}}$ User $\xrightarrow{\text{purchase}}$ Product & 6.81\% & 6.76\% & 0.0\% & 0.0\% & 0\% & 0\% & 0\% & 0\%\\
User $\xrightarrow{\text{purchase}}$ Product $\xrightarrow{\text{described}}$ Word $\xrightarrow{\text{described}}$ Product & 6.26\% & 6.22\% & 1.55\% & 1.5\% & 0\% & 0\% & 0\% & 0\%\\
User $\xrightarrow{\text{purchase}}$ Product $\xrightarrow{\text{also\_bought}}$ Product $\xrightarrow{\text{also\_bought}}$ Product & 4.51\% & 4.47\% & 6.78\% & 6.53\% & 0\% & 0\% & 0\% & 0\%\\
\rowcolor{lightcyan}
User $\xrightarrow{\text{like}^\text{\SnowflakeChevron}}$ Brand $\xrightarrow{\text{produced\_by}^\text{\SnowflakeChevron}}$ Product & 2.28\% & 2.54\% & 4.59\% & 5.59\% & 14.25\% & 18.32\% & 19.59\% & 24.49\%\\
\bottomrule
\end{tabular}
}
\caption{Proportion of the most frequent patterns used by the agents on the test set of the CDs dataset for warm users ($^\text{\color{OrangeRed}\tiny\faHotjar}$) and cold users ($^\text{\color{NavyBlue}\tiny\SnowflakeChevron}$). Rows highlighted in blue denote cold start-compatible patterns that can be used for recommending an item to a strict cold user.}
\label{table:cds_patterns}
\end{table*}

\begin{table*}[h]
\centering
\resizebox{\textwidth}{!}{
\begin{tabular}{lcccccccc}
 & \textbf{PGPR$_a^\text{\color{OrangeRed}\tiny\faHotjar}$} & \textbf{PGPR$_0^\text{\color{OrangeRed}\tiny\faHotjar}$} & \textbf{UPGPR$_a^\text{\color{OrangeRed}\tiny\faHotjar}$} & \textbf{UPGPR$_0^\text{\color{OrangeRed}\tiny\faHotjar}$} & \textbf{PGPR$_a^\text{\color{NavyBlue}\tiny\SnowflakeChevron}$} & \textbf{PGPR$_0^\text{\color{NavyBlue}\tiny\SnowflakeChevron}$} & \textbf{UPGPR$_a^\text{\color{NavyBlue}\tiny\SnowflakeChevron}$} & \textbf{UPGPR$_0^\text{\color{NavyBlue}\tiny\SnowflakeChevron}$} \\
 \midrule
User $\xrightarrow{\text{mentioned}}$ Word $\xrightarrow{\text{described}}$ Product & 65.4\% & 61.2\% & 49.6\% & 49.18\% & 0\% & 0\% & 0\% & 0\%\\
\rowcolor{lightcyan}
User $\xrightarrow{\text{interested\_in}^\text{\color{NavyBlue}\SnowflakeChevron}}$ Category $\xrightarrow{\text{belong\_to}^\text{\color{NavyBlue}\SnowflakeChevron}}$ Product & 30.19\% & 31.14\% & 29.18\% & 28.03\% & 78.96\% & 77.22\% & 70.46\% & 57.42\%\\
\rowcolor{lightcyan}
User $\xrightarrow{\text{like}^\text{\color{NavyBlue}\SnowflakeChevron}}$ Brand $\xrightarrow{\text{produced\_by}^\text{\color{NavyBlue}\SnowflakeChevron}}$ Product & 3.88\% & 4.22\% & 19.87\% & 21.45\% & 20.99\% & 22.75\% & 29.49\% & 42.53\%\\
User $\xrightarrow{\text{purchase}}$ Product $\xrightarrow{\text{also\_bought}}$ Product $\xrightarrow{\text{also\_bought}}$ Product & 0.36\% & 1.3\% & 1.18\% & 1.16\% & 0\% & 0\% & 0\% & 0\%\\
User $\xrightarrow{\text{purchase}}$ Product $\xrightarrow{\text{described}}$ Word $\xrightarrow{\text{described}}$ Product & 0.08\% & 1.89\% & 0.13\% & 0.13\% & 0\% & 0\% & 0\% & 0\%\\
User $\xrightarrow{\text{purchase}}$ Product $\xrightarrow{\text{also\_viewed}}$ Product $\xrightarrow{\text{also\_bought}}$ Product & 0.03\% & 0.06\% & 0.01\% & 0.02\% & 0\% & 0\% & 0\% & 0\%\\
\bottomrule
\end{tabular}
}
\caption{Proportion of the most frequent patterns used by the agents on the test set of the Cellphones dataset for warm users ($^\text{\color{OrangeRed}\tiny\faHotjar}$) and cold users ($^\text{\color{NavyBlue}\tiny\SnowflakeChevron}$). Rows highlighted in blue denote cold start-compatible patterns that can be used for recommending an item to a strict cold user.}
\label{table:cell_patterns}
\end{table*}

\begin{table*}[h]
\centering
\resizebox{\textwidth}{!}{
\begin{tabular}{lcccccccc}
 & \textbf{PGPR$_a^\text{\color{OrangeRed}\tiny\faHotjar}$} & \textbf{PGPR$_0^\text{\color{OrangeRed}\tiny\faHotjar}$} & \textbf{UPGPR$_a^\text{\color{OrangeRed}\tiny\faHotjar}$} & \textbf{UPGPR$_0^\text{\color{OrangeRed}\tiny\faHotjar}$} & \textbf{PGPR$_a^\text{\color{NavyBlue}\tiny\SnowflakeChevron}$} & \textbf{PGPR$_0^\text{\color{NavyBlue}\tiny\SnowflakeChevron}$} & \textbf{UPGPR$_a^\text{\color{NavyBlue}\tiny\SnowflakeChevron}$} & \textbf{UPGPR$_0^\text{\color{NavyBlue}\tiny\SnowflakeChevron}$} \\
 \midrule
User $\xrightarrow{\text{mentioned}}$ Word $\xrightarrow{\text{described}}$ Product & 74.23\% & 74.41\% & 58.27\% & 57.5\% & 0\% & 0\% & 0\% & 0\%\\
\rowcolor{lightcyan}
User $\xrightarrow{\text{interested\_in}^\text{\color{NavyBlue}\SnowflakeChevron}}$ Category $\xrightarrow{\text{belong\_to}^\text{\color{NavyBlue}\SnowflakeChevron}}$ Product & 16.24\% & 20.48\% & 37.19\% & 38.43\% & 92.02\% & 95.7\% & 81.99\% & 90.31\%\\
User $\xrightarrow{\text{purchase}}$ Product $\xrightarrow{\text{also\_bought}}$ Product $\xrightarrow{\text{also\_bought}}$ Product & 4.49\% & 2.04\% & 0.73\% & 0.73\% & 0\% & 0\% & 0\% & 0\%\\
User $\xrightarrow{\text{purchase}}$ Product $\xrightarrow{\text{described}}$ Word $\xrightarrow{\text{described}}$ Product & 4.01\% & 2.28\% & 0.13\% & 0.13\% & 0\% & 0\% & 0\% & 0\%\\
\rowcolor{lightcyan}
User $\xrightarrow{\text{like}^\text{\color{NavyBlue}\SnowflakeChevron}}$ Brand $\xrightarrow{\text{produced\_by}^\text{\color{NavyBlue}\SnowflakeChevron}}$ Product & 0.41\% & 0.26\% & 3.63\% & 3.18\% & 7.41\% & 4.14\% & 17.85\% & 9.69\%\\
\bottomrule
\end{tabular}
}
\caption{Proportion of the most frequent patterns used by the agents on the test set of the Clothing dataset for warm users ($^\text{\color{OrangeRed}\tiny\faHotjar}$) and cold users ($^\text{\color{NavyBlue}\tiny\SnowflakeChevron}$). Rows highlighted in blue denote cold start-compatible patterns that can be used for recommending an item to a strict cold user.}
\label{table:cloth_patterns}
\end{table*}

\begin{table*}[h]
\centering
\resizebox{\textwidth}{!}{%
\begin{tabular}{lcccccccc}
 & \textbf{PGPR$_a^\text{\color{OrangeRed}\tiny\faHotjar}$} & \textbf{PGPR$_0^\text{\color{OrangeRed}\tiny\faHotjar}$} & \textbf{UPGPR$_a^\text{\color{OrangeRed}\tiny\faHotjar}$} & \textbf{UPGPR$_0^\text{\color{OrangeRed}\tiny\faHotjar}$} & \textbf{PGPR$_a^\text{\color{NavyBlue}\tiny\SnowflakeChevron}$} & \textbf{PGPR$_0^\text{\color{NavyBlue}\tiny\SnowflakeChevron}$} & \textbf{UPGPR$_a^\text{\color{NavyBlue}\tiny\SnowflakeChevron}$} & \textbf{UPGPR$_0^\text{\color{NavyBlue}\tiny\SnowflakeChevron}$} \\
 \midrule
 \rowcolor{lightcyan}
User $\xrightarrow{\text{know}^\text{\color{NavyBlue}\SnowflakeChevron}}$ Skill $\xrightarrow{\text{know}^\text{\color{NavyBlue}\SnowflakeChevron}}$ User $\xrightarrow{\text{enroll}}$ Course & 59.0\% & 24.11\% & 9.86\% & 10.58\% & 65.45\% & 16.31\% & 6.5\% & 4.24\%\\
\rowcolor{lightcyan}
User $\xrightarrow{\text{know}^\text{\color{NavyBlue}\SnowflakeChevron}}$ Skill $\xrightarrow{\text{cover}^\text{\color{NavyBlue}\SnowflakeChevron}}$ Course & 40.18\% & 72.36\% & 83.84\% & 82.34\% & 34.55\% & 83.69\% & 93.5\% & 95.76\%\\
User $\xrightarrow{\text{enroll}}$ Course $\xrightarrow{\text{belong\_to}^\text{\color{NavyBlue}\SnowflakeChevron}}$ Category $\xrightarrow{\text{belong\_to}^\text{\color{NavyBlue}\SnowflakeChevron}}$ Course & 0.47\% & 1.81\% & 0.27\% & 0.24\% & 0\% & 0\% & 0\% & 0\%\\
User $\xrightarrow{\text{enroll}}$ Course $\xrightarrow{\text{taught\_by}^\text{\color{NavyBlue}\SnowflakeChevron}}$ Teacher $\xrightarrow{\text{taught\_by}^\text{\color{NavyBlue}\SnowflakeChevron}}$ Course & 0.29\% & 0.9\% & 0.3\% & 0.31\% & 0\% & 0\% & 0\% & 0\%\\
User $\xrightarrow{\text{enroll}}$ Course $\xrightarrow{\text{cover}^\text{\color{NavyBlue}\SnowflakeChevron}}$ Skill $\xrightarrow{\text{cover}^\text{\color{NavyBlue}\SnowflakeChevron}}$ Course & 0.07\% & 0.08\% & 0.12\% & 0.05\% & 0\% & 0\% & 0\% & 0\%\\
User $\xrightarrow{\text{enroll}}$ Course $\xrightarrow{\text{enroll}}$ User $\xrightarrow{\text{enroll}}$ Course & 0.0\% & 0.74\% & 5.61\% & 6.48\% & 0\% & 0\% & 0\% & 0\%\\
\bottomrule
\end{tabular}
}
\caption{Proportion of all the patterns used by the agents on the test set of the COCO dataset for warm users ($^\text{\color{OrangeRed}\tiny\faHotjar}$) and cold users ($^\text{\color{NavyBlue}\tiny\SnowflakeChevron}$). Rows highlighted in blue denote patterns that can be used for recommending an item to a strict cold user.}
\label{table:coco_patterns}
\end{table*}

\end{document}